\documentclass[12pt]{article}
\pdfoutput=1
\usepackage{amsmath}
\allowdisplaybreaks

\usepackage{natbib}
\usepackage[colorlinks=true,linkcolor=blue,citecolor=blue]{hyperref}
\usepackage{amsthm}
\usepackage{fullpage}
\usepackage{setspace}
\usepackage{subfigure}
\usepackage{multirow}
\usepackage[ruled,vlined]{algorithm2e}
\usepackage{algpseudocode}

\usepackage[table,xcdraw]{xcolor}

\usepackage{dcolumn}
\newcolumntype{d}[1]{D{.}{.}{#1}}

\newtheorem{theorem}{Theorem}

\usepackage{bm}
\usepackage{mathrsfs}
\usepackage{amsmath}
\usepackage{epsfig}
\usepackage{graphics}
\usepackage{amssymb}
\usepackage{color}
\usepackage{verbatim}

\newcommand{\bZ}{\mathbf{Z}}
\newcommand{\bv}{\mathbf{v}}
\newcommand{\bV}{\mathbf{V}}
\newcommand{\bH}{\mathbf{H}}
\newcommand{\bq}{\mathbf{q}}
\newcommand{\bbeta}{\mbox{\boldmath${\beta}$}}
\newcommand{\bSigma}{\mbox{\boldmath${\Sigma}$}}

\usepackage{authblk}

\title{Optimal subsampling for the Cox proportional hazards model with massive survival data} 

\author[2]{Nan Qiao}
\author[3]{Wangcheng Li}
\author[4]{Feng Xiao}
\author[1,2]{Cunjie Lin \thanks{Corresponding author: lincunjie@ruc.edu.cn}} 
\author[5,6]{Yong Zhou \thanks{Corresponding author: yzhou@fem.ecnu.edu.cn}} 

\affil[1]{Center for Applied Statistics, Renmin University of China, Beijing 100872, China}
\affil[2]{School of Statistics, Renmin University of China, Beijing 100872, China}
\affil[3]{School of Statistics, Beijing Normal University, Beijing 100875, China}
\affil[4]{Sichuan Rural Credit Union, Chengdu 61000, China}
\affil[5]{Key Laboratory of Advanced Theory and Application in Statistics and Data Science-MOE, Shanghai 200062,China}
\affil[6]{Academy of Statistics and Interdisciplinary Sciences, East China Normal University, Shanghai 200062, China}
\date{} 

\begin{document}
\maketitle
\begin{abstract}
The use of massive survival data has become common in survival analysis. In this study, a subsampling algorithm is proposed for the Cox proportional hazards model with time-dependent covariates when the sample is extraordinarily large but computing resources are relatively limited. A subsample estimator is developed by maximizing the weighted partial likelihood; it is shown to have consistency and asymptotic normality. By minimizing the asymptotic mean squared error of the subsample estimator, the optimal subsampling probabilities are formulated with explicit expressions. Simulation studies show that the proposed method can satisfactorily approximate the estimator of the full dataset. The proposed method is then applied to corporate loan and breast cancer datasets, with different censoring rates, and the outcomes confirm its practical advantages.

\end{abstract}



\section{Introduction}
\label{intro}

Survival analysis is widely used in biostatistics, marketing, economics, demography, and sociology. For example, the application of survival models to examine credit risk has grown rapidly over recent decades \citep{bellotti2009credit, djeundje2019dynamic}. The ability to predict the default time or risk for borrowers provides a pivotal index for loan officers to formulate policy and reduce the potential losses caused by loan defaults. With the development of data collection, exploration, and utilization, statistical analysis for massive survival data is also becoming desirable. However, some challenges remain. Indeed, although many traditional statistical methods are valid for use with finite samples, it is often computationally infeasible to adopt statistical analysis for massive data when computing resources are limited. For example, the Cox proportional hazards model \citep{cox1972}, which is widely adopted in survival analysis, is difficult to apply when using an extraordinarily large sample.

The statistical approaches in studies analyzing massive data can be categorized into two types. One strategy is to employ a divide-and-conquer (DAC) scheme. Standard DAC algorithms divide samples into $K$ subsets, construct an estimator using each subset, and combine these subset-specific estimators to form the final DAC estimator. For example, based on a linear model,  \cite{Dobriban2021} studied a weighted average estimator of parameters from different workers. In a quantile regression model, \cite{chen2020} summarized the statistics of different data blocks and approximated the estimator of the entire dataset with an asymptotically negligible approximation error. Using high-dimensional covariates, \cite{chen2014split} developed a penalized DAC estimator for generalized linear models.

DAC aims to analyze full datasets using a parallel or distributed computing platform. However, when computing resources are limited, a subsampling approach is preferred, under which the estimation is based on a small subsample drawn from the full dataset. Here, the subsample can be far smaller than the full sample. Nonetheless, despite using limited computing power, by extracting as much information from the full dataset as possible, this approach provides a practical solution to reducing the computational burden without overly reducing estimation accuracy. Most studies along this line consider a heterogeneous sampling strategy under which the data points, including more information, can be sampled with higher probabilities. For example, \cite{ma2015statistical} proposed a sketching leverage score-based subsampling strategy in a linear regression model. Using a logistic regression model,  \cite{wang2019more}, \cite{ai2021optimal} and  \cite{wang2021optimal} derived the optimal subsampling probabilities by minimizing the trace of the asymptotic variance-covariance matrix for the subsample-based estimator. Further, a deterministic method without random sampling, named ``information-based optimal subdata selection'', was proposed by  \cite{wang2019information} and  \cite{cheng2020information} for linear and logistic regression models, respectively. This method aims to find data points that have the maximal determinant of information matrix.
Moreover, \cite{zuo2021sampling} extended the optimal subsampling technique to the additive hazards model \citep{lin1994}. However, additive risk is often questioned in applications and is infrequently used, since it is hard to guarantee the non-negativity of the hazard.

Considering the limitations of the additive hazards model, this study focuses on the Cox proportional hazards model, which has been popular in survival analysis because of its lucid interpretation and satisfactory statistical properties. For instance,  \cite{dai2020} developed a distributed algorithm based on the Alternating Direction Method of Multipliers framework for the Cox proportional hazards model when massive data are stored at different institutions. 

Motivated by the appealing properties of the Cox proportional hazards model and desirable computational efficiency of subsampling approaches, we aim to develop an optimal subsampling procedure for this model using massive right-censored data. To the best of our knowledge, no subsampling procedures have thus far been developed for this model. Specifically, we derive the consistency and asymptotic distribution of the subsample estimator given the full dataset. Using the asymptotic properties, we then find the optimal subsampling probabilities that minimize the asymptotic mean squared error (AMSE). The proposed procedure contributes to the body of knowledge because the subsampling probability has an explicit expression, which makes it easy to implement in practice. More importantly, the proposed subsample estimator can approximate the full dataset estimator and reduce the computational burden. The numerical investigation presented later in the paper also supports the effectiveness of this method.

The remainder of this article is organized as follows. In Section \ref{sec2}, we review the Cox proportional hazards model and present our subsampling procedure. The asymptotic properties of the subsample estimator are also established. In Section \ref{sec3}, we evaluate the proposed method using numerical simulations. Real-world examples using corporate loan default data and breast cancer data are presented in Section \ref{sec4}. In Section \ref{sec5}, we conclude. The technical proofs are presented in the Appendix.

\section{Methods}
\label{sec2}
\subsection{Cox proportional hazards model and full dataset estimation}

Let $T$ be the event time of interest and $\boldsymbol{Z} = \{\boldsymbol{Z}(t): 0 \leq t< \infty \}$ be the $p \times 1$ vector of the covariates, which may be time-dependent. 
Consider the following Cox proportional hazards model:
\begin{eqnarray*}
\lambda(t \mid \boldsymbol{Z})= \lambda_{0}(t) \exp \left(\boldsymbol{\beta}_{0}^{\top} \boldsymbol{Z}(t)\right),
\end{eqnarray*}
where $\lambda_{0}(t)$ is the unspecified baseline hazard function and $\boldsymbol{\beta}_{0}$ is the true value of the $p$-dimensional regression coefficients. Under right censoring, the observed survival data are $X=\min(T,C)$ and $\delta=I(T\leq C)$, where $C$ is the censoring time and $\delta$ is the failure indicator. In this study, we assume that $T$ and $C$ are independent given the covariates $\boldsymbol{Z}$. Suppose that $\left\{(\boldsymbol{Z}_{i}, X_{i}, \delta_{i}), i = 1,\cdots,N\right\}$ are $N$ independent and identically distributed observations and that the censoring rate for the full dataset is $1-\delta=1-\sum_{i=1}^N\delta_i/N.$

Based on the full dataset, the regression coefficient vector $\boldsymbol{\beta}_{0}$ is usually estimated by maximizing the partial likelihood \citep{cox1975partial}, namely, 
\begin{eqnarray*}
\hat{\boldsymbol{\beta}}_{N}=\underset{\beta}{\operatorname{argmax}}~ \ell(\boldsymbol{\beta}),
\end{eqnarray*}
where $\ell(\boldsymbol{\beta})=\frac{1}{N}\sum_{i=1}^N\ell_i(\boldsymbol{\beta})$ is the log-partial likelihood function and 
\begin{eqnarray*}
\ell_{i}(\boldsymbol{\beta})= \delta_i \left[\boldsymbol{\beta}^{\top} \boldsymbol{Z}_{i}(X_i) - \log \left\{\sum_{j=1}^N I(X_j\geq X_i) \exp(\boldsymbol{\beta}^{\top} \boldsymbol{Z}_{j}(X_i) )\right\}\right].
\end{eqnarray*}
Equivalently, the estimator $\hat{\boldsymbol{\beta}}_N$ is the solution to the likelihood equation $U(\boldsymbol{\beta})=0$, where
\begin{eqnarray}\label{likelihood}
U(\boldsymbol{\beta})=\frac{1}{N} \sum_{i=1}^{N} \int_{0}^{\tau}\left\{\boldsymbol{Z}_{i}(t) -\bar{\boldsymbol{Z}}(\bbeta,t)\right\} d N_{i}(t),
\end{eqnarray}
where $\bar{\boldsymbol{Z}}(\bbeta,t)=S^{(1)}(\bbeta,t)/S^{(0)}(\bbeta,t)$, $S^{(k)}(\bbeta,t)=\frac{1}{N}\sum_{i=1}^N Y_i(t)\exp(\bbeta^\top \bZ_i (t) )\{ \bZ_i(t) \}^{\otimes k}$, for $k=0,1,2$, $N_i(t)=I(X_i\leq t, \delta_i=1)$ is the observed failure counting process and $Y_i(t)=I(X_i\geq t)$ is the at-risk indicator. Here, for the vector $\bv$, $\bv^{\otimes 0}=1$, $\bv^{\otimes 1}=\bv$, $\bv^{\otimes 2}=\bv\bv^\top$. Given the estimator $\hat{\boldsymbol{\beta}}_N$, the cumulative baseline hazard $\Lambda_0(t)=\int_0^t\lambda(s)ds$ can be estimated using the Breslow estimator, that is,
\begin{eqnarray*}
\hat{\Lambda}_{0}(t,\hat{\boldsymbol{\beta}}_N)=\int_0^t\frac{\sum_{i=1}^NdN_i(s)}{\sum_{j=1}^NY_j(s)\exp\left( \hat{\boldsymbol{\beta}}^{\top}_{N} \boldsymbol{Z}_{j}(s) \right)}.
\end{eqnarray*}

In contrast to the additive hazards model \citep{lin1994}, the likelihood equation (\ref{likelihood}) does not have a general closed-form solution. Therefore, iterative procedures such as the Newton--Raphson algorithm are often adopted to find the solution numerically. By ordering the survival times, we can use cumulative sums and differences to calculate the gradient of the likelihood function in the $O(N^2)$ computational complexity \citep{tarkhan2020}. However, when the sample $N$ is too large, the model can be computationally expensive to fit. Thus, a feasible subsampling procedure is necessary to estimate $\boldsymbol{\beta}$ and $\Lambda_0(t)$.

\subsection{Subsampling-based estimation}

This study aims to develop a subsampling-based estimator that achieves the same asymptotic efficiency as $\hat{\boldsymbol{\beta}}_{N}$ but can be computed efficiently. Specifically, we draw a random subsample of size $r (\ll N)$ from the full dataset with replacement according to the subsampling probabilities $\{\pi_i: i=1,\cdots,N\}$. Here, $\pi_i$ may depend on the full dataset and satisfy $\sum_{i \in S_{1}} \pi_{i}= \delta$ and $\sum_{i \in S_{0}} \pi_{i}= 1- \delta$, where $\delta$ is the failure rate, while $S_{0}=\left\{i: \delta_{i}=0 \right\}$ and $S_{1}=\left\{i: \delta_{i}=1 \right\}$ represent the index sets of censored and uncensored individuals, respectively. Denote the corresponding subsample as $\left\{ \left(\boldsymbol{Z}^{*}_{i}, X^{*}_{i}, \delta^{*}_{i}\right): i=1,\cdots, r\right\}$ with the subsampling probabilities $\{\pi^{*}_{i}, i=1, \ldots, r\}$. Using the subsample, 
we 
can estimate the regression coefficient $\boldsymbol{\beta}$ by maximizing the weighted partial likelihood:
\begin{eqnarray*}
\ell^*(\bbeta)=\frac{1}{Nr}\sum_{i=1}^r\frac{\delta_i^*}{\pi_i^*}\left[\bbeta^\top\bZ_i^*(X_i^*) -\log\left\{\sum_{j=1}^r\frac{1}{\pi_j^*}I(X_j^*\geq X_i^*)\exp(\bbeta^\top \bZ_j^*(X_i^*) )\right\}\right].
\end{eqnarray*}
By defining $ N^{*}_{i}(t)= I\left(X^{*}_{i} \leq t , \delta^{*}_{i}=1\right)$,  $ Y^{*}_{i}(t) $ $ = I\left(X^{*}_{i} \geq t\right)$, and $S^{(k)}_*(\bbeta,t)=\frac{1}{Nr}\sum_{i=1}^r\frac{1}{\pi_i^*}Y_i^*(t)\exp(\bbeta^\top \bZ_i^*(t) ) \{\bZ_i^* (t) \}^{\otimes k}$, for $k=0,1,2$,
this is equivalent to solving the weighted estimating function:
 \begin{eqnarray}
\label{u1}
\mathbf{U}^{*}(\boldsymbol{\beta})=\frac{1}{N r} \sum_{i=1}^{r} \frac{1}{\pi_{i}^{*}} U_{i}^{*}(\boldsymbol{\beta})=0,
\end{eqnarray}
where $U_{i}^{*}(\boldsymbol{\beta})=  \int_{0}^{\tau}\left\{\boldsymbol{Z}^{*}_{i}(t) -\bar{\boldsymbol{Z}}^{*}(\bbeta,t)\right\} d N^{*}_{i}(t)$ and $\bar{\bZ}^*(\bbeta,t)=S^{(1)}_*(\bbeta,t)/S^{(0)}_*(\bbeta,t)$.

Let $\tilde{\boldsymbol{\beta}}$ be the solution of $\mathbf{U}^{*}(\boldsymbol{\beta})=0$. Given the full dataset $\mathcal{F}_N=\{(\boldsymbol{Z}_{i}, X_{i}, \delta_{i}),i=1,\cdots,N\}$, $U^*(\bbeta)$ is asymptotically unbiased toward $U(\bbeta)$. Thus, the subsample estimator $\tilde{\boldsymbol{\beta}}$ can approximate the full dataset estimator $\hat{{\boldsymbol{\beta}}}_{N}$, as shown in Theorem \ref{th1}. 

\begin{theorem}
\label{th1}
Under the regularity conditions (C1)--(C6) in the Appendix, given $\mathcal{F}_N$, as $N \rightarrow \infty$ and $r \rightarrow \infty$, for any $\varepsilon>0$, with a probability approaching one, there exist finite $\Delta_{\varepsilon}$ and $r_{\varepsilon}$, such that 
\begin{eqnarray*}
P\left(\left\|\tilde{\boldsymbol{\beta}}-\hat{\boldsymbol{\beta}}_{N}\right\| \geq r^{-1 / 2} \Delta_{\varepsilon} \mid \mathcal{F}_N\right)<\varepsilon,
\end{eqnarray*}
for all $r\geq r_\varepsilon$.
\end{theorem}

Theorem \ref{th1} indicates that $\tilde{\boldsymbol{\beta}}$ can approximate $\hat{\boldsymbol{\beta}}_{N}$ with an error of the order of $O_{P \mid \mathcal{F}_N}(r^{-1/2})$, where the probability measure in $O_{P \mid \mathcal{F}_N}(\cdot)$ is the conditional probability given $\mathcal{F}_N$. Using $\tilde{\boldsymbol{\beta}}$, we can obtain the corresponding subsampling-based Breslow estimator for $\Lambda_0(t)$: 
\begin{eqnarray}
\label{lambda}
\tilde{\Lambda}_{0} (t, \tilde{\boldsymbol{\beta}}) = \int_0^t\frac{\sum_{i=1}^r\pi_i^{*-1}dN_i^*(s)}{\sum_{i=1}^r\pi_i^{*-1}Y_i^*(s)\exp(\tilde{\bbeta}^\top\bZ_i^*(s) )}.
\end{eqnarray}
Based on Theorem \ref{th1}, it is easy to show that for any $t\in (0,\tau)$, $\tilde{\Lambda}_{0} (t,\tilde{\boldsymbol{\beta}})- \hat{\Lambda}_{0} (t,\hat{\boldsymbol{\beta}}_N)  = O_{P \mid \mathcal{F}_{N}}\left(r^{-1 / 2}\right)$
as $N \rightarrow \infty$ and $r \rightarrow \infty$. Besides consistency, we derive the asymptotic distribution of $\tilde{\bbeta}$.
\begin{theorem}
\label{th2}
Under the regularity conditions (C1)--(C6) in the Appendix, given $\mathcal{F}_{N}$, as $N \rightarrow \infty$ and $ r \rightarrow \infty$, we have 
\begin{eqnarray*}
\sqrt{r}(\tilde{\bbeta}-\hat{\bbeta}_N)\stackrel{d}{\rightarrow} N(0,\bSigma),
\end{eqnarray*}
where $\stackrel{d}{\rightarrow}$ denotes the convergence in distribution, $\bSigma=\bH^{-1}\bV\bH^{-1}$ with
\begin{eqnarray*}
\bH&=&\frac{1}{N}\sum_{i=1}^N\int_0^\tau\left[\frac{S^{(2)}(\bbeta_0,t)}{S^{(0)}(\bbeta_0,t)}-\left(\frac{S^{(1)}(\bbeta_0,t)}{S^{(0)}(\bbeta_0,t)}\right)^{\otimes 2}\right]dN_i(t),\\
\bV&=&\frac{1}{N^2}\sum_{i=1}^N\frac{1}{\pi_i}\int_0^\tau\{\bZ_i (t)-\bar{\bZ}(\bbeta_0,t)\}^{\otimes 2}dN_i(t)+\frac{1}{N^2}\sum_{i=1}^N\frac{1}{\pi_i}\bq_i(\bbeta_0)^{\otimes 2},
\end{eqnarray*}
in which
$\bq_i(\bbeta_0)=\int_0^\tau\{\bZ_i (t)-\bar{\bZ}(\bbeta_0,t)\}\frac{Y_i(t)\exp(\bbeta_0^\top \bZ_i (t))}{S^{(0)}(\bbeta_0,t)}d\bar{N}(t)$ and 
$\bar{N}(t)=\frac{1}{N}\sum_{i=1}^{N} N_{i}(t)$.
\end{theorem}

Based on the asymptotic distribution in Theorem \ref{th2}, we can show that the subsample estimator $\tilde{\Lambda}_0(t,\tilde{\bbeta})$ also converges to a normal distribution, that is, as $N\rightarrow \infty$ and $r\rightarrow \infty$, for any $t\in [0,\tau]$,
$\sqrt{r}\left\{\tilde{\Lambda}_0(t,\tilde{\bbeta})-\hat{\Lambda}_0(t,\hat{\bbeta}_N)\right\}
\stackrel{d}{\rightarrow} N(0,\Sigma_{\Lambda}),$ where $\Sigma_{\Lambda}=\Gamma^\top\bSigma\Gamma+\Psi+\Gamma^\top\bH^{-1}\Phi$: 
\begin{eqnarray*}
\Gamma &=& \int_0^t\frac{S^{(1)}(\bbeta_0,s)}{S^{(0)}(\bbeta_0,s)^2}d\bar{N}(s),~~~~\Phi =\frac{1}{N^2}\sum_{i=1}^N\frac{1}{\pi_i}\int_0^\tau\frac{\left\{\bZ_i (t) -\bar{\bZ}(\bbeta_0,t)\right\}}{S^{(0)}(\bbeta_0,t)}dN_i(t),\\
\Psi &=& \frac{1}{N^2}\sum_{i=1}^N\frac{1}{\pi_i}\int_0^t\frac{dN_i(s)}{S^{(0)}(\bbeta_0,s)^2}+\frac{1}{N^2}\sum_{i=1}^N\frac{1}{\pi_i}\left[\int_0^t\frac{Y_i(s)\exp(\bbeta_0^\top\bZ_i (t) )}{S^{(0)}(\bbeta_0,s)^2}d\bar{N}(s)\right]^2.
\end{eqnarray*}

The proofs of the asymptotic properties of $\tilde{\bbeta}$ and $\tilde{\Lambda}_0(t,\tilde{\bbeta})$ are provided in the Appendix. From these proofs, we observe that the asymptotic variance of $\tilde{\Lambda}_0(t,\tilde{\bbeta})-\hat{\Lambda}_0(t,\hat{\bbeta}_N)$ consists of three components: $\Gamma^\top\bSigma\Gamma$ is the asymptotic variance of $\sqrt{r}\left\{\tilde{\Lambda}_0(t,\tilde{\bbeta})-\tilde{\Lambda}_0(t,\hat{\bbeta}_N)\right\}$, which is related to the asymptotic distribution of $\tilde{\bbeta}-\hat{\bbeta}_N$; $\Psi$ is the asymptotic variance of $\sqrt{r}\left\{\tilde{\Lambda}_0(t,\hat{\bbeta}_N)-\hat{\Lambda}_0(t,\hat{\bbeta}_N)\right\}$; and $\Gamma^\top\bH^{-1}\Phi$ is the covariance of these two components. Each component of $\Sigma_{\Lambda}$ depends on the subsampling probabilities $\pi_i$, making these components more complex to optimize than those of $\bSigma$. Thus, in this study, we consider the optimal subsampling probabilities based on the asymptotic variance of $\tilde{\bbeta}$.

\subsection{Subsampling strategy}
In this section, we derive the optimal subsampling probabilities to better approximate $\hat{\bbeta}_N$ using the result in Theorem \ref{th2}. Since the uniform subsampling strategy with $\pi_i=1/N$ for $i=1,\cdots, N$ may be suboptimal for approximating the parameters, we consider the optimal subsampling probabilities that minimize the AMSE of $\tilde{\bbeta}$ in approximating $\hat{\bbeta}_N$, that is, by minimizing $\mbox {AMSE}(\tilde{\bbeta})=tr(\bSigma)$, $tr(\cdot)$ is the trace of a matrix. Since $\bSigma=\bH^{-1}\bV\bH^{-1}$ is a matrix and $\bH$ is free of $\pi_i$, this is equivalent to minimizing $tr(\bV)$. A simple calculation shows that 
\begin{eqnarray*}
tr(\bV)&=&tr\left(\frac{1}{N^2}\sum_{i=1}^N\frac{1}{\pi_i}\int_0^\tau\{\bZ_i (t) -\bar{\bZ}(\bbeta_0,t)\}^{\otimes 2}dN_i(t)+\frac{1}{N^2}\sum_{i=1}^N\frac{1}{\pi_i}\bq_i(\bbeta_0)^{\otimes 2}\right)\\
&=&\frac{1}{N^2}\sum_{i\in S_0}\frac{1}{\pi_i}tr\left(\int_0^\tau\{\bZ_i (t) -\bar{\bZ}(\bbeta_0,t)\}^{\otimes 2}dN_i(t)+\bq_i(\bbeta_0)^{\otimes 2}\right)\\
&&+\frac{1}{N^2}\sum_{i\in S_1}\frac{1}{\pi_i}tr\left(\int_0^\tau\{\bZ_i (t)-\bar{\bZ}(\bbeta_0,t)\}^{\otimes 2}dN_i(t)+\bq_i(\bbeta_0)^{\otimes 2}\right).
\end{eqnarray*}
For $i\in S_0$, $dN_i(t)=0$; thus, $tr(\bV)$ reduces to 
\begin{eqnarray*}
tr(\bV)&=& \frac{1}{N^2}\sum_{i\in S_0}\frac{1}{\pi_i}tr\left(\bq_i(\bbeta_0)^{\otimes 2}\right) \\
&& +\frac{1}{N^2}\sum_{i\in S_1}\frac{1}{\pi_i}tr\left(\int_0^\tau\{\bZ_i (t) -\bar{\bZ}(\bbeta_0,t)\}^{\otimes 2}dN_i(t)+\bq_i(\bbeta_0)^{\otimes 2} \right).
\end{eqnarray*}
Based on this formulation, we can obtain the explicit expression of the optimal subsampling probabilities. 
Specifically, if the sampling probabilities $\pi_i,  i=1,\cdots ,N$ are chosen as
\begin{eqnarray}\label{pi_0}
\pi^{opt}_i=(1-\delta)\frac{tr^{1/2}\left(\bq_i(\bbeta_0)^{\otimes 2}\right)}{\sum_{i \in S_0}tr^{1/2}\left(\bq_i(\bbeta_0)^{\otimes 2}\right)},
\end{eqnarray}
for $i\in S_0,$ and 
\begin{eqnarray}\label{pi_1}
\pi^{opt}_i=\delta \frac{tr^{1/2}\left(\int_0^\tau\{\bZ_i (t)-\bar{\bZ}(\bbeta_0,t)\}^{\otimes 2}dN_i(t)+\bq_i(\bbeta_0)^{\otimes 2}\right)}{\sum_{i\in S_1}tr^{1/2}\left(\int_0^\tau\{\bZ_i (t)-\bar{\bZ}(\bbeta_0,t)\}^{\otimes 2}dN_i(t)+\bq_i(\bbeta_0)^{\otimes 2}\right)},
\end{eqnarray}
for $i \in S_1,$ then the AMSE of $\tilde{\bbeta}$, $tr(\bH^{-1}\bSigma\bH^{-1})$ attains its minimum.


From expressions (\ref{pi_0}) and (\ref{pi_1}), we see that the subsampling probabilities for the failure samples and censoring samples differ and that the optimal subsampling probabilities satisfy that $\sum_{i\in S_0}\pi_i^{opt}=1-\delta$ and $\sum_{i\in S_1}\pi_i^{opt}=\delta$, which ensures that the subsample has the same censoring rate as that of the full dataset. Accordingly, it can capture the censoring characteristics of the full dataset to the extent possible. Nonetheless, the sampling probabilities $\pi_i^{opt}$ involve the unknown parameter $\bbeta_0$, which is not directly implementable. Hence, we need a preliminary estimate $\tilde{\bbeta}_0$ to obtain the optimal sampling probabilities, and a natural approach is to use a subsample estimator with uniform sampling. The details are summarized in Algorithm 1.

\begin{algorithm} 
 \caption{Optimal subsampling algorithm}  \label{algorithm1}
 \begin{algorithmic}[1]
 \renewcommand{\algorithmicrequire}{\textbf{Input:}}
\Require  $\left(\mathbf{Z}_{i}, X_{i}, \delta_{i}\right),i=1,2,...,N$
\renewcommand{\algorithmicrequire}{\textbf{Output:}}
\Require Subsampling-based estimate $\tilde{\boldsymbol{\beta}}$, 
 \State 1. \emph{Pilot estimate}. Obtain a pilot sample set with sample size $r(\ll n)$ by uniform sampling based on censoring rate $\delta$. Then, compute the pilot estimate $\tilde{\boldsymbol{\beta}}_{0}$ according to equation (\ref{u1}) with $\pi_i = 1/N$ for $i=1,\cdots, N$; 

\State 2. \emph{Optimal subsampling probability}. Calculate the optimal subsampling probability with (\ref{pi_0}) and (\ref{pi_1}), in which $\bbeta_0$ is replaced by $\tilde{\bbeta}_0$; then, subsample with a replacement to obtain a subsample of size $r$ using $\pi_i^{opt}$, and denote the subsample set as $ \left \{ (\boldsymbol{Z}^{*}_{i}, X^{*}_{i}, \delta^{*}_{i}), i =1,\cdots,r \right \}$;  
 
\State 3. \emph{Subsampling-based estimate}. Use the Newton--Raphson method to solve (\ref{u1}) and obtain the estimate $\tilde{\boldsymbol{\beta}}$, resulting in the estimate $\tilde{\Lambda}_0(t,\tilde{\bbeta})$ using (\ref{lambda}). 
\end{algorithmic}
\end{algorithm}

In Algorithm \ref{algorithm1}, it takes time $O(r)$ to draw a random subsample of size $r$ and the pilot estimate $\tilde{\bbeta}_0$ requires time $O(r^2)$ to compute. To calculate the optimal subsampling probability, we first order the survival times with $O(N\log N)$ computational complexity \citep{sedgewick1977analysis} and then use the cumulative sums and differences to calculate the sampling probability in the $O(Np)$ time. Using the optimal subsampling probability, the subsample requires time $O(r)$ to obtain, while the final subsampling-based estimate $\tilde{\bbeta}$ requires time $O(r^2)$ to compute. Overall, the required time to implement Algorithm \ref{algorithm1} is $O(r^2+N\log N)$. As $r\ll N$, the advantages of the subsampling-based method increase since the method can lower the computational burden substantially without reducing the effectiveness and accuracy of the estimates of the full dataset.

To implement the statistical inference to run hypothesis testing 
or construct the confidence interval of the target parameter, we must estimate the asymptotic covariance matrix of the estimator. To reduce the computational burden, we propose estimating the asymptotic variance $\bSigma$ using the subsample $\left\{(\bZ_i^*, X_i^*,\delta_i^*): i=1,\cdots, r \right\}$. Specifically, a feasible estimate is $\tilde{\bSigma}=\tilde{\bH}^{-1}\tilde{\bV}\tilde{\bH}^{-1}$, where
\begin{eqnarray*}
\tilde{\bH}&=&\frac{1}{Nr}\sum_{i=1}^r\frac{1}{\pi_i^*}\int_0^\tau\left[\frac{S_*^{(2)}(\tilde{\bbeta},t)}{S_*^{(0)}(\tilde{\bbeta},t)}-\left(\frac{S_*^{(1)}(\tilde{\bbeta},t)}{S_*^{(0)}(\tilde{\bbeta},t)}\right)^{\otimes 2}\right]dN_i^*(t),\\
\tilde{\bV}&=&\frac{1}{N^2r^2}\sum_{i=1}^r\frac{1}{\pi_i^{*2}}\int_0^\tau\{\bZ_i^* (t) -\bar{\bZ}^*(\tilde{\bbeta},t)\}^{\otimes 2}dN_i^*(t). 
\end{eqnarray*}
Similarly, the asymptotic variance of $\tilde{\Lambda}_0(t,\tilde{\bbeta})$ can be estimated by $\tilde{\Sigma}_{\Lambda}=\tilde{\Gamma}^\top\tilde{\bSigma}\tilde{\Gamma}+\tilde{\Psi}+\tilde{\Gamma}^\top\tilde{\bH}^{-1}\tilde{\Phi}$, and $\tilde{\Gamma}, \tilde{\Psi}, \tilde{\Phi}$ can be obtained in a similar way.

\section{Numerical studies}
\label{sec3}
In this section, we present the results of numerical experiments conducted to assess the performance of the proposed method and compare it with that of alternative approaches. Under the proposed approach, the optimal subsampling is carried out on both the censored and failure samples with the probabilities (\ref{pi_0}) and (\ref{pi_1}), respectively. This is denoted as {\bf Full-opt}. For comparison, we consider two alternative methods:
(1) {\bf Uniform}: uniform sampling with replacement and the subsampling probabilities $\pi_i=1/N$ for $i=1,\cdots, N$; (2) {\bf Cen-opt}: uniform sampling on the failure samples with $\pi_i=1/N\delta$ and optimal sampling on the censored samples with the probability (\ref{pi_0}). This approach is similar to the method proposed by \cite{keret2020optimal} except that they retained all the failure samples, while we adopt uniform sampling on the failure samples to ensure fairness and comparability.

To generate the failure times, we consider the Cox proportional hazards model $\lambda(t)=\lambda_0(t)\exp(\bbeta_0^\top \bZ)$, where $\bbeta_0=(0.5,1,-0.3,-0.7, 0.4,0.6)^\top$ and $\bZ=(Z_1,\cdots,Z_6)^\top$. The covariates $(Z_1,Z_2,Z_3)^\top$ are generated from the multivariate normal distribution $N(0,\bSigma_0)$ with $\bSigma_0=(0.5^{\mid i-j \mid })_{i,j}$; moreover, $Z_4$ is generated from $Gamma(2,1)$, while $Z_5$ and $Z_6$ are generated from the Bernoulli distributions $B(1,0.5)$ and $B(1,0.3)$, respectively. We consider two baseline hazard functions with  $\lambda_0(t)=0.5$ and $\lambda_0(t)=t$. The censoring time $C_i$ is generated independently of the uniform distribution $U(0,c)$, where $c$ is set such that the censoring rates are 30\% or 50\%. Overall, there are four cases: 

Case 1: $\lambda_{0}(t)=0.5$ and the censoring rate is 30\%;

Case 2: $\lambda_{0}(t)=0.5$ and the censoring rate is 50\%;

Case 3: $\lambda_{0}(t)=t$ and the censoring rate is 30\%;

Case 4: $\lambda_{0}(t)=t$ and the censoring rate is 50\%. 

\begin{figure}[htbp]
\centering
\subfigure[$\lambda_0(t)=0.5$ and the censoring rate is 30\%]{
\begin{minipage}[t]{0.5\linewidth}
\centering
\includegraphics[width=6cm]{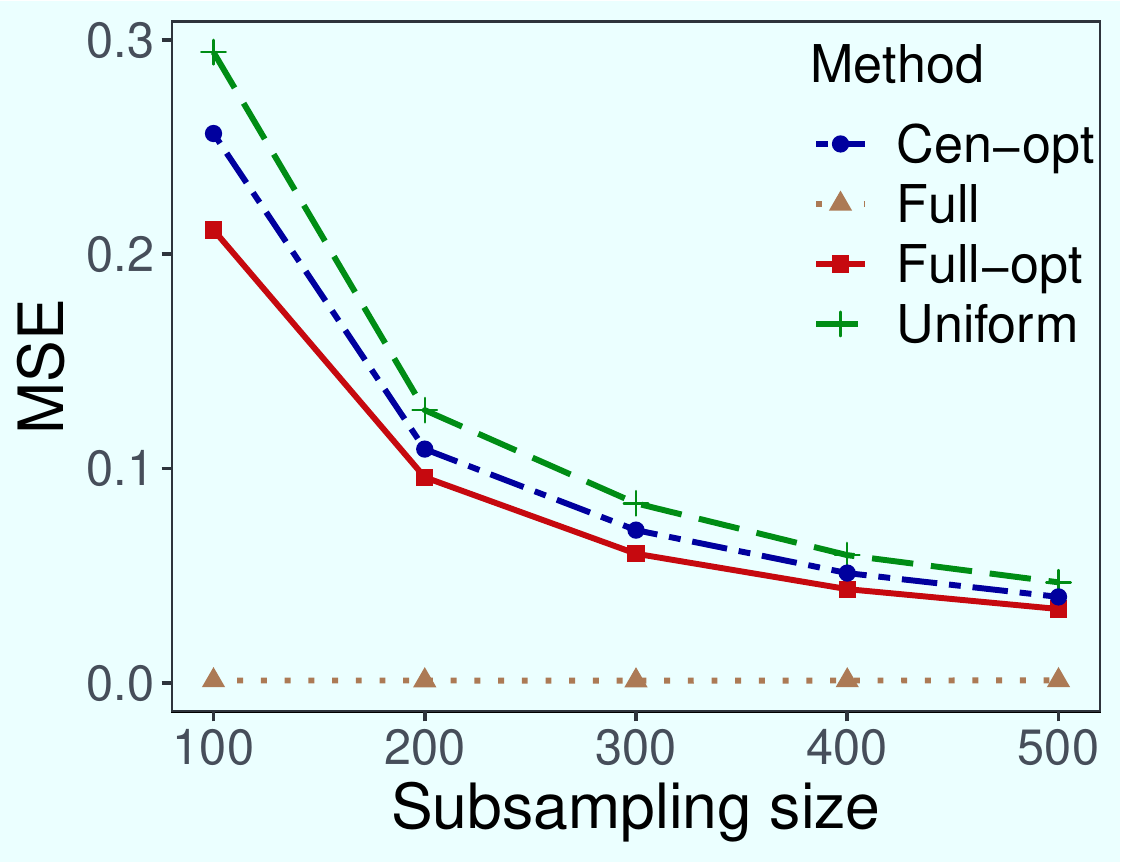}
\end{minipage}%
}%
\subfigure[$\lambda_0(t)=0.5$ and the censoring rate is 50\%]{
\begin{minipage}[t]{0.5\linewidth}
\centering
\includegraphics[width=6cm]{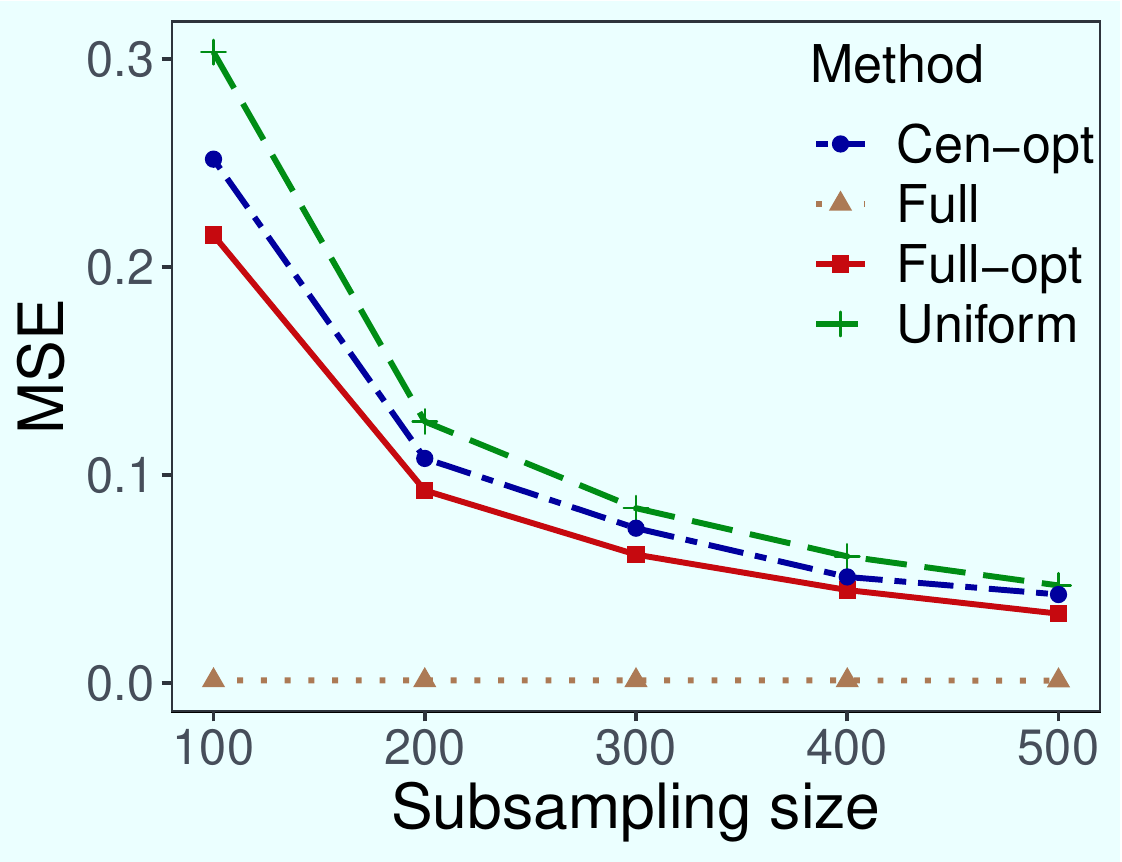}
\end{minipage}
}%

\subfigure[$\lambda_0(t)=t$ and the censoring rate is 30\%]{
\begin{minipage}[t]{0.5\linewidth}
\centering
\includegraphics[width=6cm]{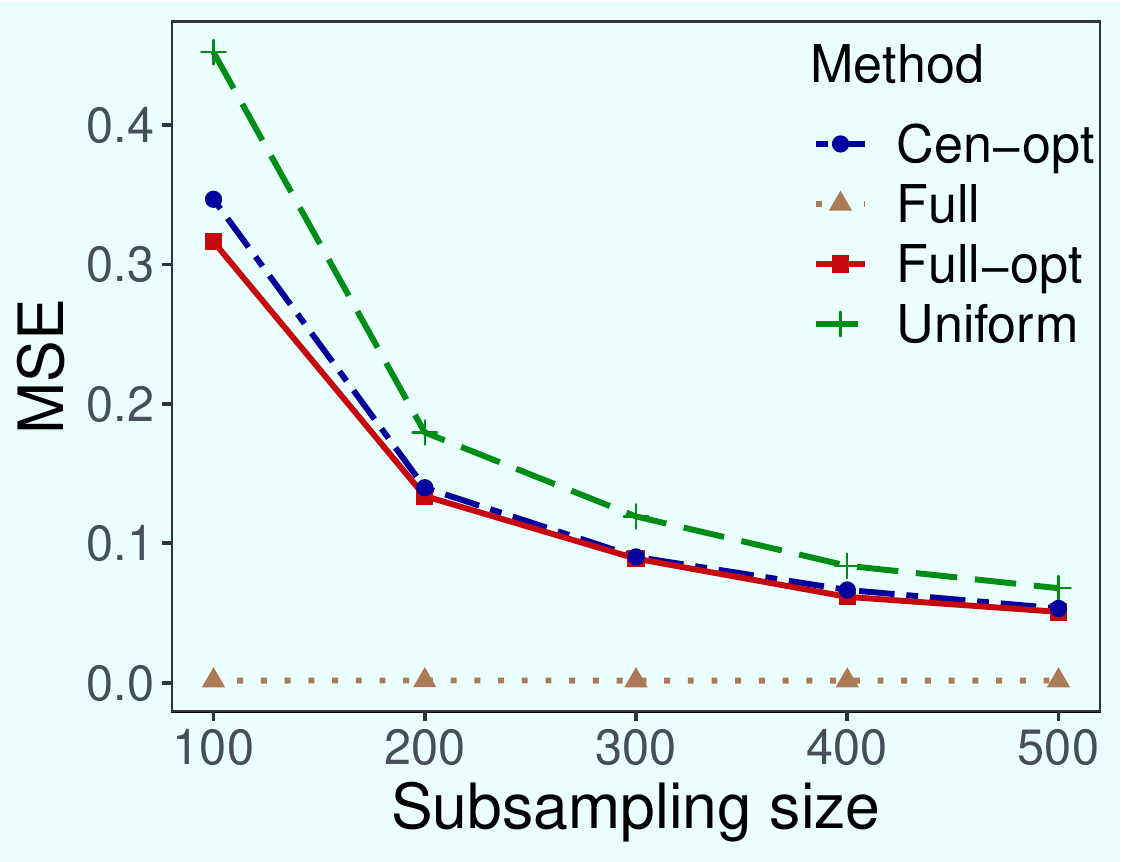}
\end{minipage}%
}%
\subfigure[$\lambda_0(t)=t$ and the censoring rate is 50\%]{
\begin{minipage}[t]{0.5\linewidth}
\centering
\includegraphics[width=6cm]{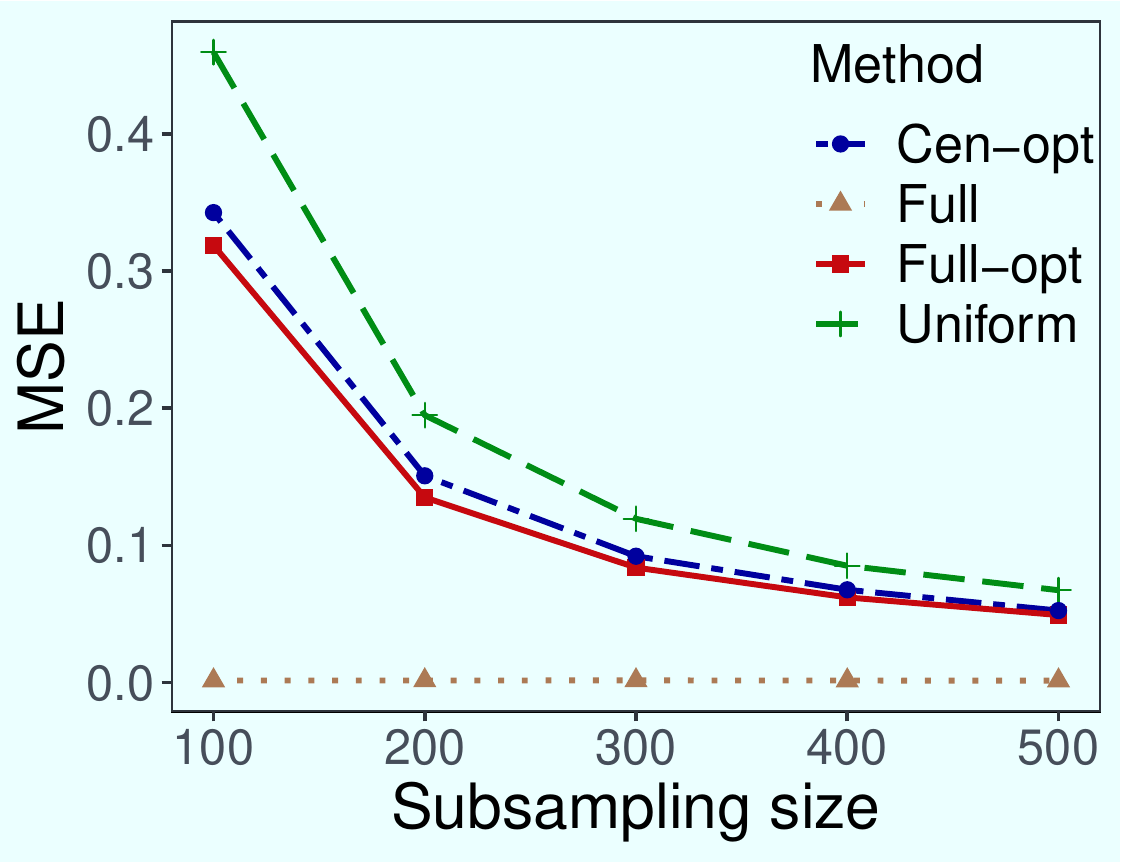}
\end{minipage}
}%
\centering
\caption{Simulation results for the MSE.}
\label{figure1}
\end{figure}

In each setting, $B=1000$ replicates are simulated using the full dataset of size $N=2\times 10^4$ and the subsample sizes $r=100, 200, 300, 400, 500$. To evaluate the performances of these different methods, for each parameter, we consider the bias (Bias) of the estimate, estimated standard error (ESE), sampling standard error (SSE), and empirical 95\% coverage probability (CP). To better compare the performance of the methods, the cumulative mean square error (MSE) is computed: $\mbox{MSE}=\frac{1}{B}\sum_{b=1}^B\|\tilde{\bbeta}^{(b)}-\bbeta_0\|^2,$ 
where $\tilde{\bbeta}^{(b)}$ is the estimate in the $b$-th replicate. Tables \ref{table.case1}--\ref{table.case4} and Figure \ref{figure1} summarize the numerical results, from which the following conclusions can be obtained.

Overall, the bias of all three estimators is small, which means that they are consistent for the true parameter. Meanwhile, the proposed method (Full-opt) usually yields a smaller bias than the alternatives, which falls as the subsample size $r$ increases from 100 to 500. Full-opt also reduces the asymptotic variance compared with the other methods, since it produces the lowest standard error in most of the situations. The uniform subsampling method tends to produce the highest standard error; this means that the optimal sampling does help improve the efficiency of the estimates. In addition, the proposed method of estimating the asymptotic variance is reliable, as the ESE is close to the SSE. Further, the confidence intervals have CPs that deviate slightly from the 0.95 nominal level. Figure \ref{figure1} shows a clear downward trend in the MSE as the subsample size increases for all three subsampling methods. Full-opt therefore has major advantages over the other two subsampling-based methods and Cen-opt performs better than Uniform in terms of the MSE, which also indicates the superiority of subsampling using the optimal probabilities.

\begin{table*}[htbp]
  \centering
 \caption{Simulation results: $\lambda_0(t)=0.5$ and the censoring rate is 30\%.}
\label{table.case1}
\resizebox{\columnwidth}{!}{
\begin{tabular}{c|c|ccc|ccc|ccc|ccc }
\hline
\multirow{3}{*}{}                                                                       & \multicolumn{1}{c|}{\multirow{3}{*}{$r$}} & \multicolumn{3}{c|}{BIAS}                                                                                                                     & \multicolumn{3}{c|}{SSE}                                                                                                                      & \multicolumn{3}{c|}{ESE}                                                                                                                      & \multicolumn{3}{c}{CP}                                                                                                                       \\
\hline
                                                                                                              & \multicolumn{1}{c|}{}                     & \multirow{2}{*}{\begin{tabular}[c]{@{}c@{}}Uni-\\ form\end{tabular}} & \multirow{2}{*}{\begin{tabular}[c]{@{}c@{}}Full-\\ opt\end{tabular}} & \multirow{2}{*}{\begin{tabular}[c]{@{}c@{}}Cen-\\ opt\end{tabular}} & \multirow{2}{*}{\begin{tabular}[c]{@{}c@{}}Uni-\\ form\end{tabular}} & \multirow{2}{*}{\begin{tabular}[c]{@{}c@{}}Full-\\ opt\end{tabular}} & \multirow{2}{*}{\begin{tabular}[c]{@{}c@{}}Cen-\\ opt\end{tabular}} & \multirow{2}{*}{\begin{tabular}[c]{@{}c@{}}Uni-\\ form\end{tabular}} & \multirow{2}{*}{\begin{tabular}[c]{@{}c@{}}Full-\\ opt\end{tabular}} & \multirow{2}{*}{\begin{tabular}[c]{@{}c@{}}Cen-\\ opt\end{tabular}} &  \multirow{2}{*}{\begin{tabular}[c]{@{}c@{}}Uni-\\ form\end{tabular}} & \multirow{2}{*}{\begin{tabular}[c]{@{}c@{}}Full-\\ opt\end{tabular}} & \multirow{2}{*}{\begin{tabular}[c]{@{}c@{}}Cen-\\ opt\end{tabular}} \\
  & \multicolumn{1}{c|}{}                     &                                                                      &                                                                      &                                                                     &                                                                      &                                                                      &                                                                     &                                                                      &                                                                      &                                                                     &                                                                      &                                                                      &                                                                     \\ \hline  
\multirow{5}{*}{$\boldsymbol{\beta}_{1}$}   & 100   & 0.024 & 0.013 & 0.027 & 0.168 & 0.137 & 0.147 & 0.160 & 0.133 & 0.142 & 0.946 & 0.937 & 0.939 \\
          & 200   & 0.009 & 0.001 & 0.013 & 0.108 & 0.088 & 0.103 & 0.108 & 0.089 & 0.096 & 0.952 & 0.956 & 0.933 \\
          & 300   & 0.007 & -0.001 & 0.012 & 0.088 & 0.072 & 0.082 & 0.087 & 0.071 & 0.078 & 0.950 & 0.942 & 0.948 \\
          & 400   & 0.008 & 0.003 & 0.004 & 0.075 & 0.064 & 0.072 & 0.075 & 0.062 & 0.068 & 0.948 & 0.947 & 0.941 \\
          & 500   & 0.006 & 0.003 & 0.004 & 0.069 & 0.056 & 0.060 & 0.066 & 0.055 & 0.060 & 0.944 & 0.939 & 0.940 \\
 \hline  
\multirow{5}{*}{$\boldsymbol{\beta}_{2}$} & 100   & 0.053 & 0.005 & 0.041 & 0.203 & 0.176 & 0.203 & 0.197 & 0.171 & 0.179 & 0.950 & 0.943 & 0.919 \\
          & 200   & 0.025 & 0.013 & 0.016 & 0.140 & 0.117 & 0.130 & 0.133 & 0.115 & 0.121 & 0.940 & 0.945 & 0.933 \\
          & 300   & 0.021 & 0.001 & 0.016 & 0.111 & 0.096 & 0.102 & 0.107 & 0.093 & 0.099 & 0.950 & 0.947 & 0.939 \\
          & 400   & 0.012 & 0.002 & 0.010 & 0.095 & 0.081 & 0.091 & 0.091 & 0.080 & 0.085 & 0.938 & 0.950 & 0.934 \\
          & 500   & 0.008 & -0.001 & 0.009 & 0.083 & 0.071 & 0.081 & 0.081 & 0.071 & 0.076 & 0.943 & 0.949 & 0.936 \\
\hline  
\multirow{5}{*}{$\boldsymbol{\beta}_{3}$} & 100   & -0.013 & -0.007 & -0.003 & 0.165 & 0.128 & 0.149 & 0.155 & 0.127 & 0.135 & 0.931 & 0.945 & 0.926 \\
          & 200   & 0.000 & -0.004 & -0.003 & 0.107 & 0.088 & 0.101 & 0.105 & 0.086 & 0.092 & 0.938 & 0.940 & 0.921 \\
          & 300   & -0.006 & 0.000 & 0.000 & 0.085 & 0.069 & 0.078 & 0.084 & 0.069 & 0.075 & 0.957 & 0.946 & 0.929 \\
          & 400   & -0.003 & 0.000 & -0.005 & 0.074 & 0.062 & 0.071 & 0.072 & 0.059 & 0.065 & 0.950 & 0.937 & 0.931 \\
          & 500   & -0.002 & -0.001 & -0.005 & 0.065 & 0.054 & 0.058 & 0.064 & 0.052 & 0.058 & 0.947 & 0.947 & 0.951 \\
\hline  
\multirow{5}{*}{$\boldsymbol{\beta}_{4}$} &  100   & -0.038 & -0.010 & -0.034 & 0.145 & 0.117 & 0.133 & 0.133 & 0.112 & 0.122 & 0.934 & 0.938 & 0.925 \\
          & 200   & -0.022 & -0.003 & -0.017 & 0.094 & 0.076 & 0.085 & 0.090 & 0.075 & 0.083 & 0.940 & 0.957 & 0.940 \\
          & 300   & -0.015 & 0.001 & -0.011 & 0.073 & 0.061 & 0.070 & 0.072 & 0.060 & 0.066 & 0.951 & 0.948 & 0.939 \\
          & 400   & -0.007 & -0.004 & -0.008 & 0.066 & 0.053 & 0.058 & 0.062 & 0.052 & 0.057 & 0.943 & 0.945 & 0.944 \\
          & 500   & -0.006 & -0.002 & -0.009 & 0.057 & 0.047 & 0.051 & 0.055 & 0.046 & 0.051 & 0.941 & 0.939 & 0.954 \\
\hline  
\multirow{5}{*}{$\boldsymbol{\beta}_{5}$} & 100   & 0.019 & 0.000 & 0.023 & 0.277 & 0.251 & 0.256 & 0.262 & 0.232 & 0.236 & 0.940 & 0.940 & 0.931 \\
          & 200   & 0.015 & -0.003 & 0.009 & 0.176 & 0.158 & 0.169 & 0.178 & 0.158 & 0.162 & 0.956 & 0.942 & 0.944 \\
          & 300   & -0.002 & -0.001 & 0.008 & 0.143 & 0.130 & 0.137 & 0.143 & 0.127 & 0.130 & 0.954 & 0.952 & 0.944 \\
          & 400   & 0.007 & -0.003 & 0.002 & 0.121 & 0.106 & 0.111 & 0.123 & 0.110 & 0.113 & 0.957 & 0.957 & 0.958 \\
          & 500   & 0.008 & 0.005 & 0.000 & 0.112 & 0.096 & 0.101 & 0.110 & 0.098 & 0.101 & 0.943 & 0.952 & 0.951 \\
\hline  
\multirow{5}{*}{$\boldsymbol{\beta}_{6}$} & 100   & 0.022 & 0.006 & 0.025 & 0.307 & 0.262 & 0.289 & 0.297 & 0.261 & 0.265 & 0.946 & 0.955 & 0.929 \\
          & 200   & 0.010 & 0.010 & 0.004 & 0.208 & 0.189 & 0.187 & 0.202 & 0.179 & 0.182 & 0.948 & 0.937 & 0.945 \\
          & 300   & 0.003 & 0.009 & 0.011 & 0.173 & 0.143 & 0.155 & 0.163 & 0.144 & 0.148 & 0.945 & 0.953 & 0.944 \\
          & 400   & -0.006 & -0.001 & 0.003 & 0.142 & 0.124 & 0.130 & 0.139 & 0.123 & 0.127 & 0.947 & 0.945 & 0.952 \\
          & 500   & 0.001 & 0.004 & 0.007 & 0.122 & 0.109 & 0.117 & 0.124 & 0.110 & 0.114 & 0.955 & 0.951 & 0.943 \\
\hline
    \end{tabular}}%
\end{table*}%

\begin{table*}[htbp]
  \centering
 \caption{Simulation results: $\lambda_0(t)=0.5$ and the censoring rate is 50\%.}
\label{table.case2}
\resizebox{\columnwidth}{!}{
\begin{tabular}{c|c|ccc|ccc|ccc|ccc }
\hline
\multirow{3}{*}{}                                                                       & \multicolumn{1}{c|}{\multirow{3}{*}{$r$}} & \multicolumn{3}{c|}{BIAS}                                                                                                                     & \multicolumn{3}{c|}{SSE}                                                                                                                      & \multicolumn{3}{c|}{ESE}                                                                                                                      & \multicolumn{3}{c}{CP}                                                                                                                       \\
\hline
                                                                                                              & \multicolumn{1}{c|}{}                     & \multirow{2}{*}{\begin{tabular}[c]{@{}c@{}}Uni-\\ form\end{tabular}} & \multirow{2}{*}{\begin{tabular}[c]{@{}c@{}}Full-\\ opt\end{tabular}} & \multirow{2}{*}{\begin{tabular}[c]{@{}c@{}}Cen-\\ opt\end{tabular}} & \multirow{2}{*}{\begin{tabular}[c]{@{}c@{}}Uni-\\ form\end{tabular}} & \multirow{2}{*}{\begin{tabular}[c]{@{}c@{}}Full-\\ opt\end{tabular}} & \multirow{2}{*}{\begin{tabular}[c]{@{}c@{}}Cen-\\ opt\end{tabular}} & \multirow{2}{*}{\begin{tabular}[c]{@{}c@{}}Uni-\\ form\end{tabular}} & \multirow{2}{*}{\begin{tabular}[c]{@{}c@{}}Full-\\ opt\end{tabular}} & \multirow{2}{*}{\begin{tabular}[c]{@{}c@{}}Cen-\\ opt\end{tabular}} &  \multirow{2}{*}{\begin{tabular}[c]{@{}c@{}}Uni-\\ form\end{tabular}} & \multirow{2}{*}{\begin{tabular}[c]{@{}c@{}}Full-\\ opt\end{tabular}} & \multirow{2}{*}{\begin{tabular}[c]{@{}c@{}}Cen-\\ opt\end{tabular}} \\
  & \multicolumn{1}{c|}{}                     &                                                                      &                                                                      &                                                                     &                                                                      &                                                                      &                                                                     &                                                                      &                                                                      &                                                                     &                                                                      &                                                                      &                                                                     \\ \hline  
\multirow{5}{*}{$\boldsymbol{\beta}_{1}$}   & 100   & 0.033 & 0.011 & 0.028 & 0.175 & 0.137 & 0.155 & 0.163 & 0.132 & 0.143 & 0.933 & 0.936 & 0.926 \\
          & 200   & 0.010 & 0.003 & 0.012 & 0.107 & 0.089 & 0.101 & 0.108 & 0.088 & 0.097 & 0.954 & 0.954 & 0.939 \\
          & 300   & 0.009 & 0.003 & 0.007 & 0.090 & 0.072 & 0.085 & 0.087 & 0.071 & 0.078 & 0.951 & 0.952 & 0.926 \\
          & 400   & 0.012 & 0.000 & 0.009 & 0.079 & 0.063 & 0.070 & 0.075 & 0.061 & 0.068 & 0.931 & 0.949 & 0.944 \\
          & 500   & 0.005 & 0.000 & 0.004 & 0.065 & 0.053 & 0.062 & 0.067 & 0.054 & 0.060 & 0.953 & 0.957 & 0.944 \\
 \hline  
\multirow{5}{*}{$\boldsymbol{\beta}_{2}$} & 100   & 0.053 & 0.026 & 0.047 & 0.207 & 0.182 & 0.196 & 0.200 & 0.170 & 0.181 & 0.933 & 0.939 & 0.927 \\
          & 200   & 0.028 & 0.008 & 0.017 & 0.134 & 0.117 & 0.131 & 0.134 & 0.115 & 0.122 & 0.944 & 0.950 & 0.935 \\
          & 300   & 0.018 & 0.003 & 0.022 & 0.109 & 0.094 & 0.107 & 0.107 & 0.092 & 0.099 & 0.941 & 0.947 & 0.928 \\
          & 400   & 0.009 & 0.000 & 0.006 & 0.095 & 0.081 & 0.088 & 0.092 & 0.079 & 0.085 & 0.938 & 0.943 & 0.947 \\
          & 500   & 0.010 & 0.000 & 0.013 & 0.084 & 0.072 & 0.082 & 0.082 & 0.070 & 0.076 & 0.946 & 0.952 & 0.925 \\
\hline  
\multirow{5}{*}{$\boldsymbol{\beta}_{3}$} & 100   & -0.017 & -0.010 & -0.014 & 0.166 & 0.138 & 0.157 & 0.157 & 0.126 & 0.138 & 0.941 & 0.935 & 0.917 \\
          & 200   & -0.011 & -0.002 & -0.004 & 0.109 & 0.083 & 0.099 & 0.105 & 0.084 & 0.093 & 0.943 & 0.959 & 0.935 \\
          & 300   & -0.004 & -0.001 & -0.005 & 0.090 & 0.066 & 0.078 & 0.084 & 0.068 & 0.076 & 0.932 & 0.956 & 0.942 \\
          & 400   & -0.005 & 0.001 & 0.000 & 0.074 & 0.060 & 0.070 & 0.072 & 0.058 & 0.065 & 0.947 & 0.951 & 0.933 \\
          & 500   & 0.000 & 0.002 & -0.001 & 0.067 & 0.052 & 0.062 & 0.064 & 0.051 & 0.058 & 0.936 & 0.953 & 0.937 \\
\hline  
\multirow{5}{*}{$\boldsymbol{\beta}_{4}$} &  100  & -0.035 & -0.015 & -0.034 & 0.138 & 0.115 & 0.122 & 0.132 & 0.108 & 0.119 & 0.934 & 0.932 & 0.950 \\
          & 200   & -0.021 & -0.006 & -0.014 & 0.089 & 0.072 & 0.083 & 0.088 & 0.072 & 0.080 & 0.951 & 0.959 & 0.940 \\
          & 300   & -0.011 & -0.001 & -0.015 & 0.075 & 0.057 & 0.069 & 0.071 & 0.058 & 0.065 & 0.931 & 0.949 & 0.938 \\
          & 400   & -0.010 & 0.000 & -0.007 & 0.060 & 0.051 & 0.058 & 0.061 & 0.050 & 0.056 & 0.949 & 0.944 & 0.945 \\
          & 500   & -0.012 & -0.002 & -0.008 & 0.055 & 0.046 & 0.053 & 0.054 & 0.045 & 0.050 & 0.947 & 0.949 & 0.939  \\
\hline  
\multirow{5}{*}{$\boldsymbol{\beta}_{5}$} & 100  & 0.037 & -0.002 & 0.008 & 0.272 & 0.240 & 0.255 & 0.265 & 0.232 & 0.236 & 0.939 & 0.938 & 0.928 \\
          & 200   & 0.004 & 0.006 & 0.006 & 0.181 & 0.162 & 0.160 & 0.179 & 0.157 & 0.161 & 0.948 & 0.940 & 0.959 \\
          & 300   & 0.010 & -0.003 & 0.004 & 0.149 & 0.132 & 0.137 & 0.144 & 0.126 & 0.131 & 0.944 & 0.928 & 0.940 \\
          & 400   & 0.007 & -0.003 & 0.004 & 0.129 & 0.112 & 0.112 & 0.124 & 0.108 & 0.113 & 0.942 & 0.938 & 0.951 \\
          & 500   & 0.002 & 0.000 & 0.001 & 0.110 & 0.099 & 0.104 & 0.110 & 0.096 & 0.101 & 0.943 & 0.938 & 0.946 \\
\hline  
\multirow{5}{*}{$\boldsymbol{\beta}_{6}$} & 100  & 0.034 & 0.012 & 0.034 & 0.318 & 0.269 & 0.282 & 0.299 & 0.258 & 0.267 & 0.929 & 0.939 & 0.935 \\
          & 200   & 0.012 & -0.003 & 0.003 & 0.205 & 0.181 & 0.194 & 0.202 & 0.175 & 0.181 & 0.947 & 0.942 & 0.937 \\
          & 300   & 0.005 & 0.004 & 0.013 & 0.166 & 0.150 & 0.159 & 0.162 & 0.140 & 0.147 & 0.949 & 0.934 & 0.931 \\
          & 400   & 0.015 & 0.002 & 0.003 & 0.138 & 0.124 & 0.131 & 0.139 & 0.120 & 0.126 & 0.957 & 0.944 & 0.942 \\
          & 500   & 0.005 & 0.003 & 0.003 & 0.125 & 0.103 & 0.118 & 0.124 & 0.107 & 0.113 & 0.949 & 0.953 & 0.936\\
\hline
    \end{tabular}}%
\end{table*}%

\begin{table*}[htbp]
  \centering
 \caption{Simulation results: $\lambda_0(t)=t$ and the censoring rate is 30\%.}
\label{table.case3}
\resizebox{\columnwidth}{!}{
\begin{tabular}{c|c|ccc|ccc|ccc|ccc }
\hline
\multirow{3}{*}{}                                                                       & \multicolumn{1}{c|}{\multirow{3}{*}{$r$}} & \multicolumn{3}{c|}{BIAS}                                                                                                                     & \multicolumn{3}{c|}{SSE}                                                                                                                      & \multicolumn{3}{c|}{ESE}                                                                                                                      & \multicolumn{3}{c}{CP}                                                                                                                       \\
\hline
                                                                                                              & \multicolumn{1}{c|}{}                     & \multirow{2}{*}{\begin{tabular}[c]{@{}c@{}}Uni-\\ form\end{tabular}} & \multirow{2}{*}{\begin{tabular}[c]{@{}c@{}}Full-\\ opt\end{tabular}} & \multirow{2}{*}{\begin{tabular}[c]{@{}c@{}}Cen-\\ opt\end{tabular}} & \multirow{2}{*}{\begin{tabular}[c]{@{}c@{}}Uni-\\ form\end{tabular}} & \multirow{2}{*}{\begin{tabular}[c]{@{}c@{}}Full-\\ opt\end{tabular}} & \multirow{2}{*}{\begin{tabular}[c]{@{}c@{}}Cen-\\ opt\end{tabular}} & \multirow{2}{*}{\begin{tabular}[c]{@{}c@{}}Uni-\\ form\end{tabular}} & \multirow{2}{*}{\begin{tabular}[c]{@{}c@{}}Full-\\ opt\end{tabular}} & \multirow{2}{*}{\begin{tabular}[c]{@{}c@{}}Cen-\\ opt\end{tabular}} &  \multirow{2}{*}{\begin{tabular}[c]{@{}c@{}}Uni-\\ form\end{tabular}} & \multirow{2}{*}{\begin{tabular}[c]{@{}c@{}}Full-\\ opt\end{tabular}} & \multirow{2}{*}{\begin{tabular}[c]{@{}c@{}}Cen-\\ opt\end{tabular}} \\
  & \multicolumn{1}{c|}{}                     &                                                                      &                                                                      &                                                                     &                                                                      &                                                                      &                                                                     &                                                                      &                                                                      &                                                                     &                                                                      &                                                                      &                                                                     \\ \hline  
\multirow{5}{*}{$\boldsymbol{\beta}_{1}$}   & 100 & 0.046 & 0.002 & 0.026 & 0.210 & 0.165 & 0.174 & 0.193 & 0.157 & 0.163 & 0.935 & 0.942 & 0.926 \\
          & 200 & 0.010 & 0.003 & 0.008 & 0.135 & 0.106 & 0.112 & 0.127 & 0.104 & 0.109 & 0.938 & 0.949 & 0.939 \\
          & 300 & 0.008 & 0.001 & 0.007 & 0.107 & 0.084 & 0.090 & 0.102 & 0.084 & 0.088 & 0.943 & 0.957 & 0.957 \\
          & 400 & 0.011 & -0.002 & 0.008 & 0.090 & 0.071 & 0.077 & 0.088 & 0.072 & 0.076 & 0.937 & 0.948 & 0.951 \\
          & 500 & 0.003 & 0.005 & 0.003 & 0.080 & 0.067 & 0.069 & 0.078 & 0.064 & 0.068 & 0.943 & 0.946 & 0.946 \\
 \hline  
\multirow{5}{*}{$\boldsymbol{\beta}_{2}$} & 100   & 0.071 & 0.015 & 0.039 & 0.249 & 0.210 & 0.219 & 0.233 & 0.199 & 0.204 & 0.930 & 0.934 & 0.934 \\
          & 200 & 0.029 & 0.012 & 0.015 & 0.157 & 0.132 & 0.147 & 0.154 & 0.133 & 0.138 & 0.950 & 0.961 & 0.932 \\
          & 300 & 0.021 & 0.002 & 0.013 & 0.132 & 0.113 & 0.115 & 0.123 & 0.106 & 0.110 & 0.938 & 0.932 & 0.938 \\
          & 400 & 0.012 & 0.007 & 0.005 & 0.107 & 0.091 & 0.096 & 0.106 & 0.091 & 0.095 & 0.946 & 0.951 & 0.945 \\
          & 500 & 0.014 & 0.003 & 0.014 & 0.096 & 0.082 & 0.087 & 0.094 & 0.081 & 0.085 & 0.952 & 0.942 & 0.936 \\
\hline  
\multirow{5}{*}{$\boldsymbol{\beta}_{3}$} & 100  & -0.016 & -0.008 & -0.016 & 0.196 & 0.155 & 0.175 & 0.186 & 0.151 & 0.157 & 0.949 & 0.949 & 0.932 \\
          & 200 & -0.009 & -0.005 & -0.002 & 0.128 & 0.104 & 0.108 & 0.124 & 0.101 & 0.106 & 0.939 & 0.949 & 0.934 \\
          & 300 & -0.002 & 0.000 & -0.004 & 0.100 & 0.083 & 0.086 & 0.099 & 0.080 & 0.085 & 0.951 & 0.945 & 0.951 \\
          & 400 & -0.004 & -0.003 & 0.002 & 0.083 & 0.070 & 0.075 & 0.085 & 0.069 & 0.073 & 0.954 & 0.949 & 0.937 \\
          & 500 & -0.003 & -0.002 & -0.007 & 0.078 & 0.061 & 0.067 & 0.076 & 0.061 & 0.065 & 0.945 & 0.956 & 0.948 \\
\hline  
\multirow{5}{*}{$\boldsymbol{\beta}_{4}$} &  100   & -0.045 & -0.012 & -0.035 & 0.177 & 0.151 & 0.164 & 0.164 & 0.139 & 0.148 & 0.951 & 0.931 & 0.931 \\
          & 200 & -0.019 & -0.008 & -0.012 & 0.112 & 0.089 & 0.101 & 0.110 & 0.092 & 0.098 & 0.948 & 0.955 & 0.938 \\
          & 300 & -0.014 & 0.000 & -0.013 & 0.092 & 0.076 & 0.080 & 0.089 & 0.073 & 0.080 & 0.941 & 0.942 & 0.950 \\
          & 400 & -0.011 & -0.001 & -0.007 & 0.078 & 0.062 & 0.070 & 0.076 & 0.063 & 0.068 & 0.945 & 0.956 & 0.952 \\
          & 500 & -0.006 & -0.004 & -0.008 & 0.069 & 0.056 & 0.065 & 0.067 & 0.056 & 0.061 & 0.944 & 0.955 & 0.932 \\
\hline  
\multirow{5}{*}{$\boldsymbol{\beta}_{5}$} & 100   & 0.041 & 0.015 & 0.029 & 0.330 & 0.292 & 0.288 & 0.314 & 0.277 & 0.274 & 0.937 & 0.939 & 0.945 \\
          & 200 & 0.000 & 0.004 & 0.008 & 0.216 & 0.191 & 0.192 & 0.210 & 0.187 & 0.185 & 0.950 & 0.952 & 0.937 \\
          & 300 & 0.015 & 0.003 & 0.003 & 0.174 & 0.149 & 0.150 & 0.170 & 0.151 & 0.149 & 0.944 & 0.960 & 0.948 \\
          & 400 & 0.001 & 0.000 & 0.000 & 0.144 & 0.131 & 0.132 & 0.146 & 0.129 & 0.128 & 0.962 & 0.949 & 0.941 \\
          & 500 & 0.000 & 0.000 & 0.013 & 0.133 & 0.117 & 0.115 & 0.130 & 0.115 & 0.114 & 0.941 & 0.952 & 0.941 \\
\hline  
\multirow{5}{*}{$\boldsymbol{\beta}_{6}$} & 100 & 0.039 & 0.016 & 0.027 & 0.393 & 0.335 & 0.351 & 0.364 & 0.322 & 0.317 & 0.930 & 0.939 & 0.930 \\
          & 200 & 0.014 & -0.010 & 0.022 & 0.245 & 0.222 & 0.214 & 0.243 & 0.214 & 0.213 & 0.957 & 0.949 & 0.948 \\
          & 300 & -0.002 & -0.004 & 0.015 & 0.202 & 0.185 & 0.179 & 0.195 & 0.173 & 0.171 & 0.948 & 0.931 & 0.935 \\
          & 400 & 0.002 & -0.006 & 0.010 & 0.174 & 0.149 & 0.153 & 0.168 & 0.148 & 0.148 & 0.937 & 0.944 & 0.945 \\
          & 500 & 0.004 & 0.002 & 0.005 & 0.153 & 0.138 & 0.136 & 0.150 & 0.133 & 0.132 & 0.950 & 0.938 & 0.940 \\
\hline
    \end{tabular}}%
\end{table*}%

\begin{table*}[htbp]
  \centering
 \caption{Simulation results: $\lambda_0(t)=t$ and the censoring rate is 50\%.}
\label{table.case4}
\resizebox{\columnwidth}{!}{
\begin{tabular}{c|c|ccc|ccc|ccc|ccc }
\hline
\multirow{3}{*}{}                                                                       & \multicolumn{1}{c|}{\multirow{3}{*}{$r$}} & \multicolumn{3}{c|}{BIAS}                                                                                                                     & \multicolumn{3}{c|}{SSE}                                                                                                                      & \multicolumn{3}{c|}{ESE}                                                                                                                      & \multicolumn{3}{c}{CP}                                                                                                                       \\
\hline
                                                                                                              & \multicolumn{1}{c|}{}                     & \multirow{2}{*}{\begin{tabular}[c]{@{}c@{}}Uni-\\ form\end{tabular}} & \multirow{2}{*}{\begin{tabular}[c]{@{}c@{}}Full-\\ opt\end{tabular}} & \multirow{2}{*}{\begin{tabular}[c]{@{}c@{}}Cen-\\ opt\end{tabular}} & \multirow{2}{*}{\begin{tabular}[c]{@{}c@{}}Uni-\\ form\end{tabular}} & \multirow{2}{*}{\begin{tabular}[c]{@{}c@{}}Full-\\ opt\end{tabular}} & \multirow{2}{*}{\begin{tabular}[c]{@{}c@{}}Cen-\\ opt\end{tabular}} & \multirow{2}{*}{\begin{tabular}[c]{@{}c@{}}Uni-\\ form\end{tabular}} & \multirow{2}{*}{\begin{tabular}[c]{@{}c@{}}Full-\\ opt\end{tabular}} & \multirow{2}{*}{\begin{tabular}[c]{@{}c@{}}Cen-\\ opt\end{tabular}} &  \multirow{2}{*}{\begin{tabular}[c]{@{}c@{}}Uni-\\ form\end{tabular}} & \multirow{2}{*}{\begin{tabular}[c]{@{}c@{}}Full-\\ opt\end{tabular}} & \multirow{2}{*}{\begin{tabular}[c]{@{}c@{}}Cen-\\ opt\end{tabular}} \\
  & \multicolumn{1}{c|}{}                     &                                                                      &                                                                      &                                                                     &                                                                      &                                                                      &                                                                     &                                                                      &                                                                      &                                                                     &                                                                      &                                                                      &                                                                     \\ \hline  
\multirow{5}{*}{$\boldsymbol{\beta}_{1}$}   & 100   & 0.043 & 0.013 & 0.033 & 0.210 & 0.158 & 0.176 & 0.193 & 0.156 & 0.163 & 0.938 & 0.946 & 0.926 \\
          & 200   & 0.021 & -0.004 & 0.010 & 0.135 & 0.106 & 0.111 & 0.129 & 0.103 & 0.109 & 0.943 & 0.943 & 0.955 \\
          & 300   & 0.017 & 0.007 & 0.008 & 0.106 & 0.087 & 0.094 & 0.103 & 0.083 & 0.088 & 0.953 & 0.939 & 0.941 \\
          & 400   & 0.009 & 0.004 & 0.008 & 0.090 & 0.070 & 0.081 & 0.088 & 0.071 & 0.076 & 0.948 & 0.949 & 0.932 \\
          & 500   & 0.008 & 0.003 & 0.008 & 0.079 & 0.068 & 0.068 & 0.078 & 0.063 & 0.068 & 0.953 & 0.925 & 0.949  \\
 \hline  
\multirow{5}{*}{$\boldsymbol{\beta}_{2}$} & 100    & 0.065 & 0.011 & 0.057 & 0.249 & 0.215 & 0.224 & 0.235 & 0.196 & 0.205 & 0.939 & 0.926 & 0.920 \\
          & 200   & 0.034 & 0.010 & 0.034 & 0.170 & 0.140 & 0.151 & 0.157 & 0.132 & 0.139 & 0.929 & 0.936 & 0.925 \\
          & 300   & 0.023 & 0.012 & 0.023 & 0.128 & 0.109 & 0.115 & 0.125 & 0.106 & 0.111 & 0.940 & 0.948 & 0.939 \\
          & 400   & 0.022 & 0.006 & 0.014 & 0.110 & 0.092 & 0.099 & 0.107 & 0.091 & 0.095 & 0.931 & 0.951 & 0.934 \\
          & 500   & 0.015 & 0.003 & 0.008 & 0.095 & 0.080 & 0.089 & 0.096 & 0.081 & 0.085 & 0.951 & 0.949 & 0.933 \\
\hline  
\multirow{5}{*}{$\boldsymbol{\beta}_{3}$} & 100  & -0.020 & -0.011 & -0.023 & 0.202 & 0.157 & 0.168 & 0.188 & 0.149 & 0.157 & 0.940 & 0.944 & 0.932 \\
          & 200   & -0.013 & 0.001 & -0.006 & 0.135 & 0.102 & 0.114 & 0.125 & 0.099 & 0.106 & 0.931 & 0.948 & 0.936 \\
          & 300   & -0.008 & -0.009 & -0.007 & 0.105 & 0.084 & 0.089 & 0.100 & 0.080 & 0.085 & 0.946 & 0.942 & 0.934 \\
          & 400   & -0.007 & 0.003 & -0.004 & 0.087 & 0.070 & 0.076 & 0.085 & 0.068 & 0.073 & 0.947 & 0.940 & 0.939 \\
          & 500   & -0.006 & -0.001 & -0.003 & 0.077 & 0.062 & 0.067 & 0.076 & 0.061 & 0.065 & 0.947 & 0.946 & 0.933 \\
\hline  
\multirow{5}{*}{$\boldsymbol{\beta}_{4}$} &  100  & -0.054 & -0.017 & -0.037 & 0.185 & 0.145 & 0.163 & 0.165 & 0.133 & 0.143 & 0.930 & 0.927 & 0.911 \\
          & 200   & -0.033 & 0.001 & -0.021 & 0.113 & 0.091 & 0.105 & 0.110 & 0.088 & 0.097 & 0.949 & 0.937 & 0.911 \\
          & 300   & -0.019 & -0.006 & -0.015 & 0.088 & 0.072 & 0.081 & 0.088 & 0.071 & 0.078 & 0.944 & 0.954 & 0.941 \\
          & 400   & -0.017 & -0.006 & -0.010 & 0.075 & 0.061 & 0.070 & 0.075 & 0.061 & 0.066 & 0.950 & 0.946 & 0.940 \\
          & 500   & -0.012 & -0.004 & -0.008 & 0.067 & 0.057 & 0.062 & 0.067 & 0.054 & 0.059 & 0.948 & 0.937 & 0.934  \\
\hline  
\multirow{5}{*}{$\boldsymbol{\beta}_{5}$} & 100  & 0.029 & 0.009 & 0.026 & 0.344 & 0.296 & 0.302 & 0.317 & 0.275 & 0.272 & 0.943 & 0.928 & 0.923 \\
          & 200   & 0.013 & 0.011 & 0.015 & 0.222 & 0.190 & 0.199 & 0.213 & 0.184 & 0.184 & 0.933 & 0.942 & 0.930 \\
          & 300   & 0.017 & 0.014 & 0.015 & 0.177 & 0.150 & 0.159 & 0.171 & 0.148 & 0.148 & 0.939 & 0.938 & 0.932 \\
          & 400   & 0.012 & -0.007 & 0.011 & 0.150 & 0.131 & 0.132 & 0.147 & 0.128 & 0.128 & 0.949 & 0.943 & 0.947 \\
          & 500   & 0.007 & 0.002 & 0.007 & 0.131 & 0.113 & 0.117 & 0.130 & 0.113 & 0.114 & 0.941 & 0.948 & 0.938 \\
\hline  
\multirow{5}{*}{$\boldsymbol{\beta}_{6}$} & 100  & 0.031 & 0.007 & 0.029 & 0.387 & 0.337 & 0.329 & 0.365 & 0.312 & 0.310 & 0.937 & 0.938 & 0.930 \\
          & 200   & 0.023 & 0.011 & 0.013 & 0.254 & 0.222 & 0.223 & 0.244 & 0.208 & 0.209 & 0.942 & 0.930 & 0.932 \\
          & 300   & 0.010 & 0.007 & 0.023 & 0.200 & 0.171 & 0.170 & 0.195 & 0.168 & 0.169 & 0.947 & 0.949 & 0.949 \\
          & 400   & 0.009 & -0.001 & 0.014 & 0.168 & 0.151 & 0.150 & 0.167 & 0.144 & 0.146 & 0.947 & 0.938 & 0.940 \\
          & 500   & 0.008 & 0.004 & 0.004 & 0.155 & 0.136 & 0.133 & 0.149 & 0.129 & 0.130 & 0.934 & 0.931 & 0.944 \\
\hline
    \end{tabular}}%
\end{table*}%

%
%
%

\begin{figure}[h]
\centering
\includegraphics[scale=0.35]{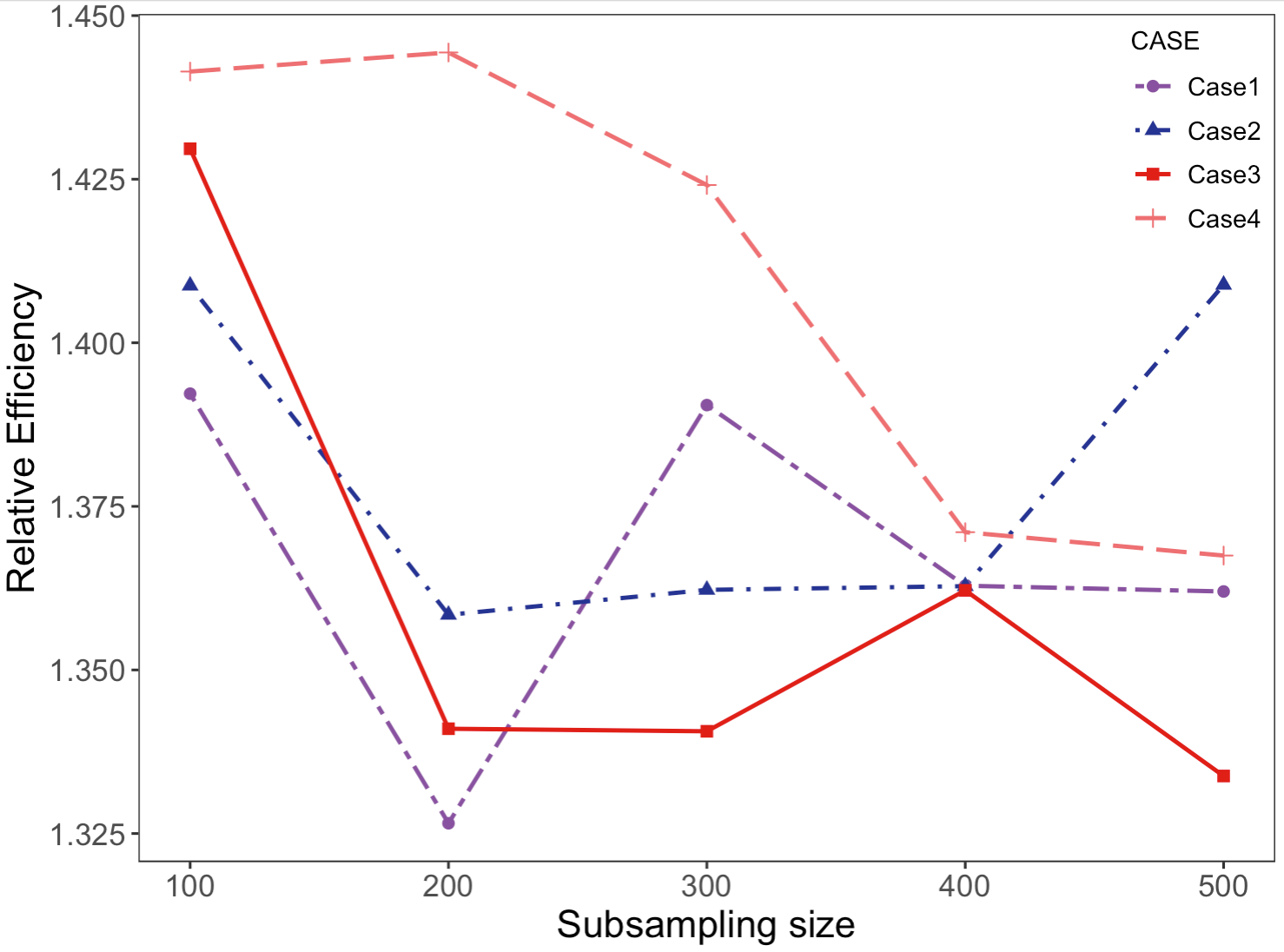}
\caption{Simulation results for relative efficiency: $\frac{\mbox{MSE}(\tilde{\bbeta}_{Unifrom})}{\mbox{MSE}(\tilde{\bbeta}_{Full-opt})}$}.
\label{re}
\end{figure}

To further evaluate the performance of Full-opt and Uniform, we show their relative efficiency, which is calculated by dividing the MSE of Uniform by the MSE of Full-opt. The result in Figure \ref{re} illustrates that the relative efficiency is above 1 for all the settings and that the values are higher when the censoring rate is 50\% rather than 30\% in most of the settings for the different subsamples. Hence, Full-opt is more efficient than Uniform, especially for a high censoring rate.

\section{Data analysis}
\label{sec4}
\subsection{Corporate loan dataset analysis}
In this subsection, we illustrate the proposed method by analyzing a corporate loan dataset from a commercial bank in southwest China. This dataset contains 27,034 loan records of companies from July 2013 to April 2021. The bank's loan service is mainly provided to the agriculture, forestry, animal husbandry and fishery, manufacturing, and wholesale and retail sales industries in China. Because most industries need a long collection and profit cycle and are somewhat likely to default and fail to repay on schedule, commercial banks must assess the default risk of a company to reduce bad debt. This is also crucial to help loan officers decide whether to approve loans \citep{yuan2018,zhang2020}.

Here, bad debt or default usually means that the principal or interest was not paid on time. Thus, we define the failure event as the default of the loan recipient; records with non-default loans over the study period are censored, and the rate of failure is 13.51\%. We are interested in the relationship between the hazard of default and the following covariates: loan amount ($Z_1$), loan rate ($Z_2$), and industry of the loan recipient. We introduce three dummy variables for the industry classification \citep{Agrawal2019}: $Z_3=1$ represents the agriculture, forestry, animal husbandry, and fishery industries; $Z_4=1$ represents the manufacturing industry; and $Z_5=1$ represents the wholesale and retail sales industry. The remaining industries such as mining, housing, and construction are classified as the baseline sectors. Thus, we consider the Cox proportional hazards model with five covariates $Z_1,\cdots, Z_5$. 

For the subsampling-based estimates, we consider Uniform, Cen-opt, and Full-opt with a subsample size $r=500$. The subsampling process is repeated
$B=1000$ times. We calculate the means and SSEs of the estimates as well as the mean of the ESEs. To gain further insights into the analysis results, we also calculate the estimates of the full dataset $\hat{\bbeta}_N=(\hat{\beta}_1,\cdots,\hat{\beta}_5)^\top$ and corresponding standard errors using all the observations. For each parameter, we calculate the MSE for the three subsampling methods. Here, $\mbox{MSE}(\tilde{\beta}_j)=\frac{1}{B}\sum_{b=1}^B(\tilde{\beta}_j^{(b)}-\hat{\beta}_{j})^2$, for $j=1,\cdots,5$. Table \ref{table5} summarizes the results.

The estimate of $\bbeta_1$ is negative, while that of $\bbeta_2$ is positive for all the methods, which is consistent with the real-world data. Indeed, the loan amount and loan rate are assigned according to the firm's credit history and credit rating. Hence, companies with excellent credit and repayment ability usually receive their requested loan amounts and obtain lower loan rates, meaning that their default risks could be lower. In the real world, companies in different industries have different profit cycles and default risks. The estimates of $\bbeta_3$, $\bbeta_4$, and $\bbeta_5$ indicate that the default risks of the $Z_3$, $Z_4$, and $Z_5$ industries are all larger than those of the other industries. Moreover, all three subsampling-based methods yield similar estimates to those derived from the full dataset. No method clearly dominates between Full-opt and Cen-opt in terms of the SSE, ESE, and MSE because these approaches share the same subsampling probability on the censoring samples and the censoring rate is nearly 86.5\%. In addition, both perform better than Uniform in terms of the SSE, ESE, and MSE, which means that the optimal subsampling method is more efficient than Uniform at approximating the estimates from the full dataset.

\begin{table*}[htbp]
  \centering
  \caption{Results of the corporate lending data analysis.}
    \begin{tabular}{c|c|c|c|c|c|c}
   \hline
          &       & $\boldsymbol{\beta}_{1}$ & $\boldsymbol{\beta}_{2}$ & $\boldsymbol{\beta}_{3}$ & $\boldsymbol{\beta}_{4}$ & $\boldsymbol{\beta}_{5}$ \\
  \hline
    \multirow{2}[2]{*}{Full dataset} & Mean   & -0.114 & 0.298 & 0.291 & 0.657 & 0.824 \\
          & SE    & 0.020 & 0.015 & 0.047 & 0.045 & 0.042 \\
   \hline
    \multirow{4}[2]{*}{Uniform} & Mean   & -0.133 & 0.308 & 0.269 & 0.653 & 0.824 \\
          & SSE    & 0.149 & 0.123 & 0.368 & 0.364 & 0.324 \\
          & ESE   & 0.152 & 0.115 & 0.360 & 0.345 & 0.317 \\
          & MSE   & 0.022 & 0.015 & 0.136 & 0.132 & 0.105 \\
    \hline
    \multirow{4}[2]{*}{Full-opt} & Mean   & -0.121 & 0.296 & 0.266 & 0.653 & 0.823 \\
          & SSE    & 0.119 & 0.101 & 0.364 & 0.348 & 0.326 \\
          & ESE   & 0.115 & 0.102 & 0.361 & 0.352 & 0.323 \\
          & MSE   & 0.014 & 0.010 & 0.133 & 0.121 & 0.106 \\
   \hline
    \multirow{4}[2]{*}{Cen-opt} & Mean   & -0.137 & 0.291 & 0.242 & 0.619 & 0.812 \\
          & SSE    & 0.139 & 0.113 & 0.358 & 0.308 & 0.300 \\
          & ESE   & 0.126 & 0.109 & 0.340 & 0.324 & 0.295 \\
          & MSE   & 0.020 & 0.013 & 0.130 & 0.096 & 0.090 \\
    \hline
    \end{tabular}%
  \label{table5}%
\end{table*}%


\subsection{Breast cancer dataset analysis}
In this subsection, we use a breast cancer dataset to further illustrate the application of the proposed subsampling method in biomedical research. Breast cancer is one of the most common malignant tumors in women. Owing to population aging and the extension of women's average life expectancy, the incidence rate and mortality of breast cancer in women aged above 65 are significantly higher than the others. Since patients of different ages have vastly different physiques and health statuses, there is no standardized treatment to meet their needs. Therefore, it is crucial to study the factors that influence the risk of being diagnosed with breast cancer. In this section, we analyze data from the Surveillance, Epidemiology, and End Results program of the National Cancer Institute. The medical records contain 49,358 female breast cancer patients from 1994 to 2003. The database collects information on each patient's survival time and related covariates such as age, tumor size, tumor stage, marital status, and estrogen/progesterone receptor status.

The median survival time of the recorded cases is 119 months and the longest observed survival time is 263 months. Owing to the sufficiently long observation period, the censoring rate is only 5.27\%. Following \citep{Rosenberg2005}, five variables are included in the Cox proportional hazards model: patient age ($Z_1$), tumor size ($Z_2$), marital status ($Z_3=1$ represents divorced patients and $Z_4=1$ represents married patients), estrogen receptor status ($Z_5=1$ represents a negative estrogen receptor and $Z_5=0$ otherwise), progesterone receptor status ($Z_6=1$ represents a negative progesterone receptor and $Z_6=0$ otherwise), and tumor stage. According to the TNM stage of breast cancer in the American Joint Committee on Cancer (Sixth Edition), tumor stage can be categorized as I, IB, IIB, IIIA, IIIB, IV, and other. We set $Z_7=1$ for tumor stage I, $Z_8 =1$ for tumor stage IB, $Z_9 =1$ for tumor stage IIB, $Z_{10} =1$ for tumor stage IIIA, $Z_{11} =1$ for tumor stage IIIB, and $Z_{12}=1$ for tumor stage IV. Regarding the continuous variables, tumor size is missing for 11.92\% of cases and the missing values are replaced with the median before fitting the Cox proportional hazards model with $Z_1,\cdots, Z_{12}$.

\begin{figure}[htbp]
\centering
\subfigure[Uniform subsampling]{
\begin{minipage}[t]{0.32\linewidth}
\centering
\includegraphics[width=4cm]{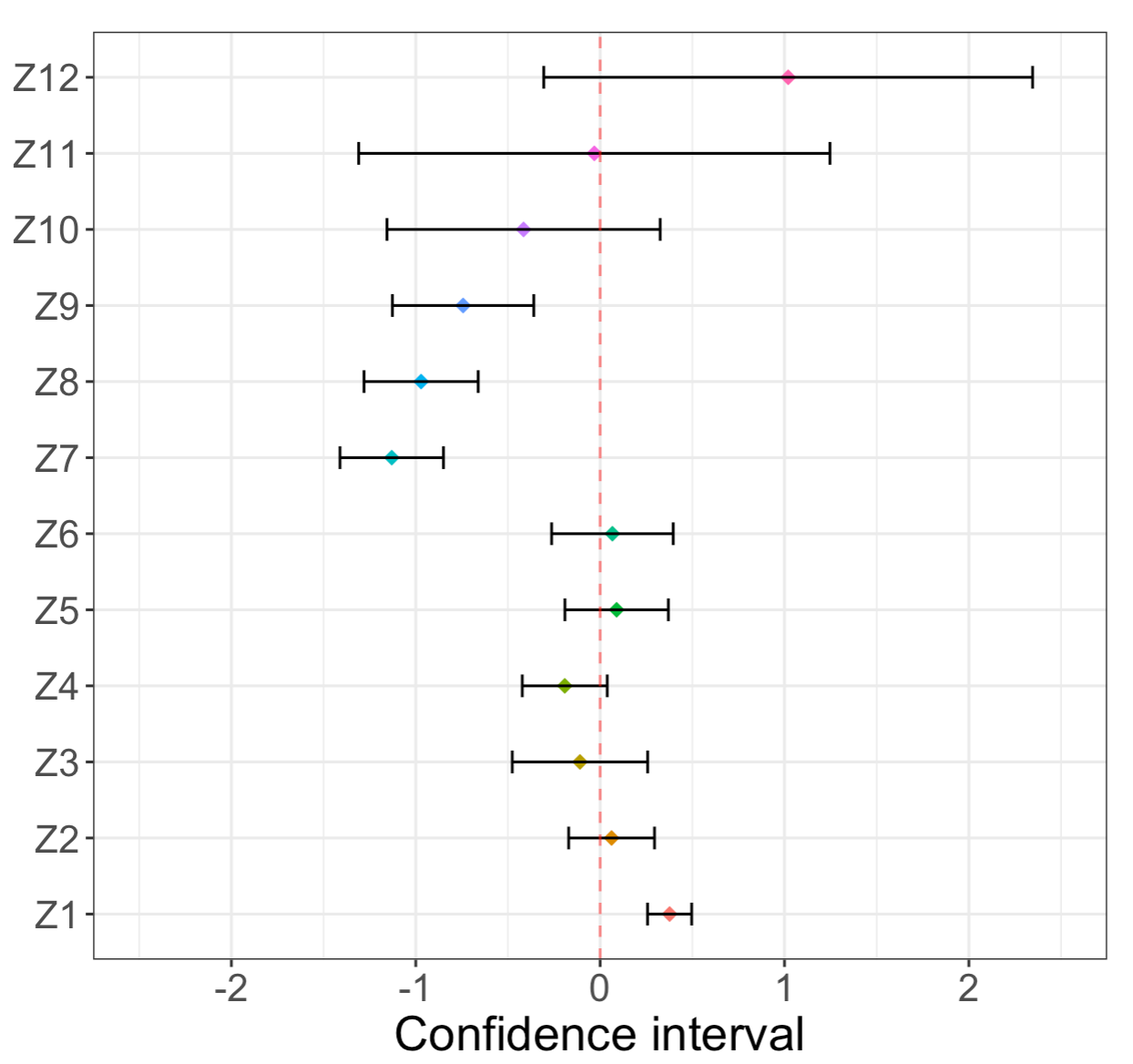}
\end{minipage}%
}%
\subfigure[Cen-opt subsampling]{
\begin{minipage}[t]{0.32\linewidth}
\centering
\includegraphics[width=4cm]{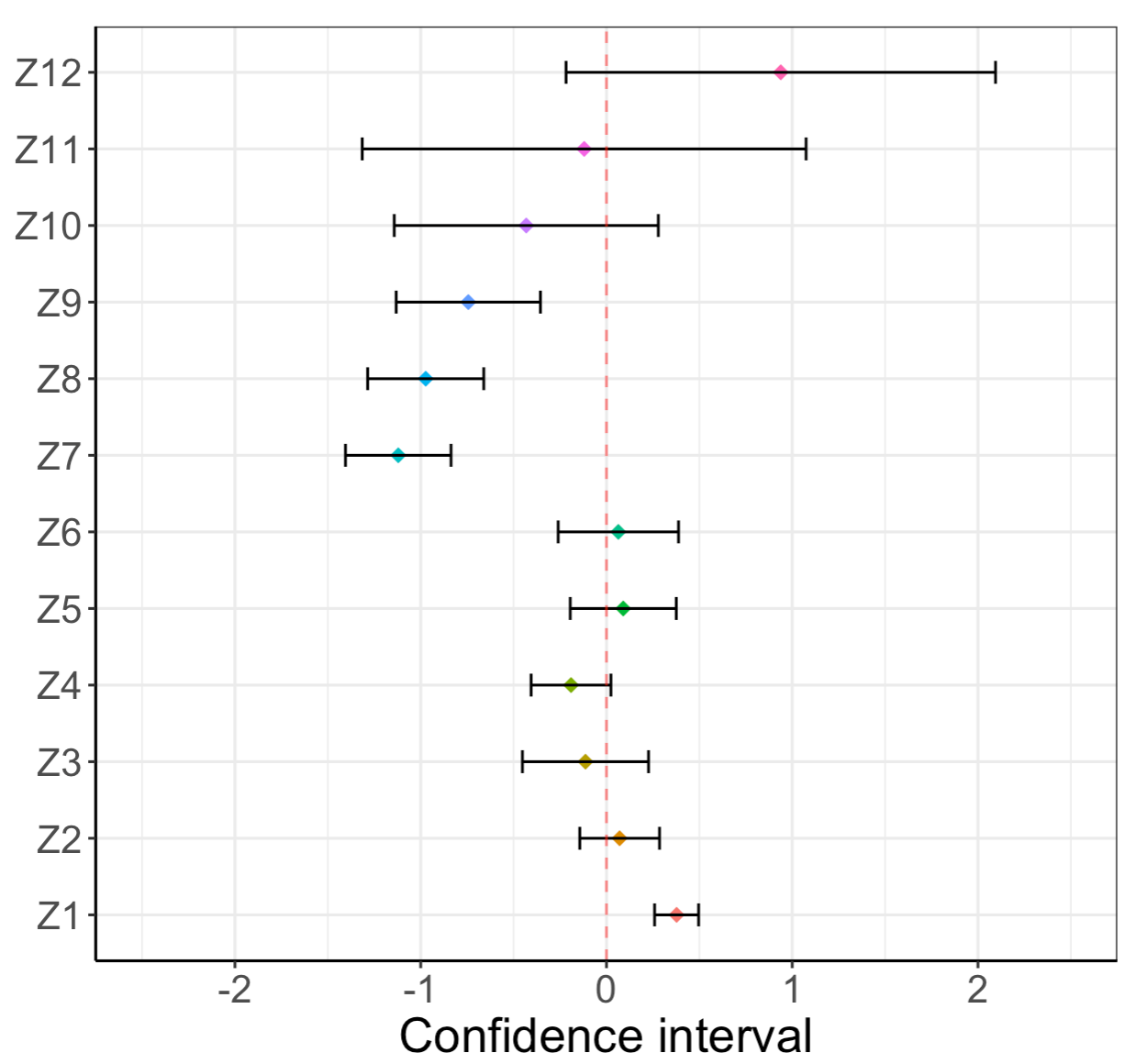}
\end{minipage}
}%
\subfigure[Full-opt subsampling]{
\begin{minipage}[t]{0.32\linewidth}
\centering
\includegraphics[width=4cm]{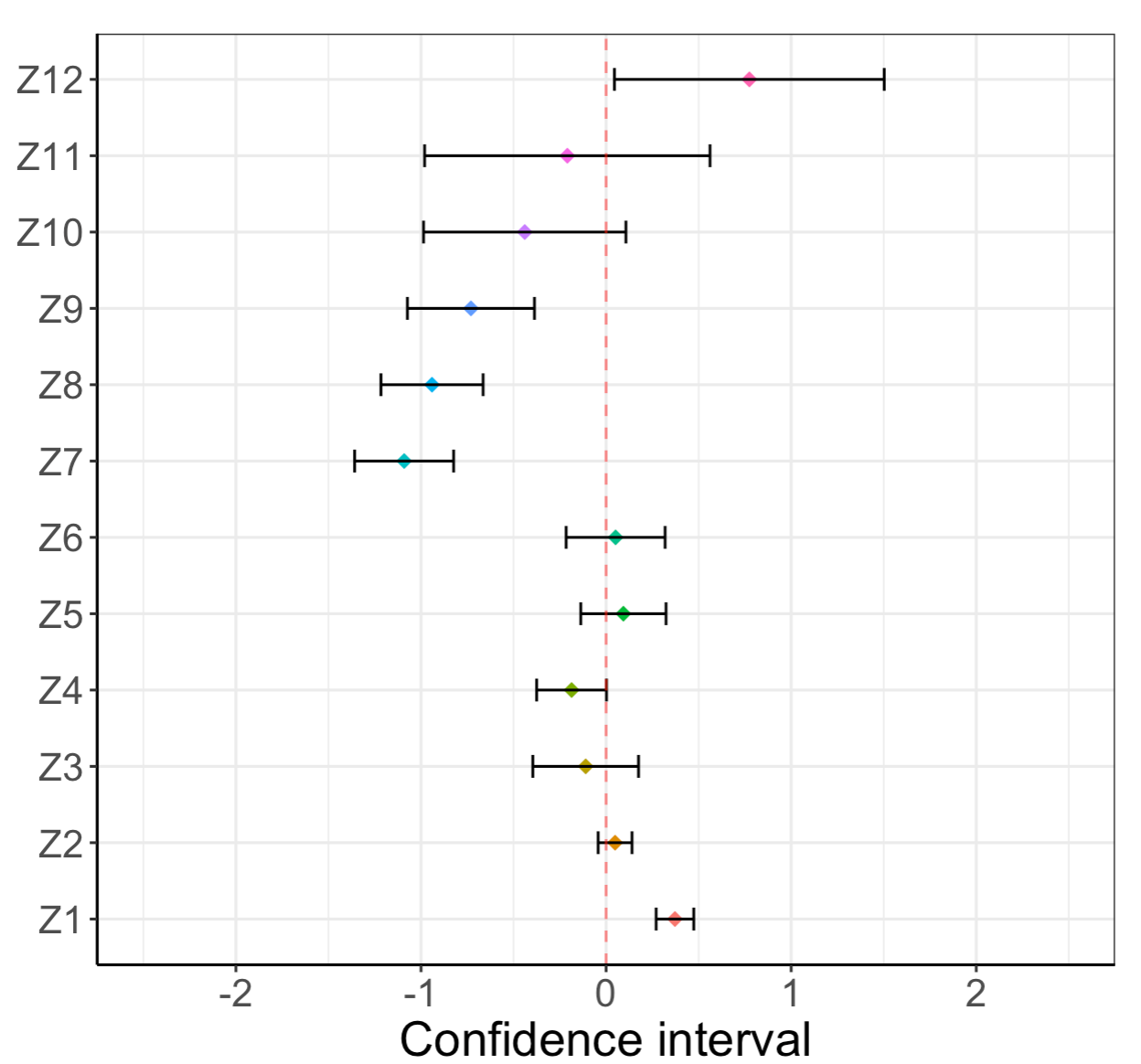}
\end{minipage}%
}%
\centering
\caption{Breast cancer dataset analysis: Confidence interval of the coefficients.}
\label{ci}
\end{figure}

%
%
Figure \ref{ci} shows the results of the coefficient estimates and confidence intervals. Full-opt yields the shortest confidence intervals of the three subsampling methods, and there are more significant variables than for Uniform and Cen-opt. Compared with the analysis of the corporate loan dataset, the advantages of Full-opt over Cen-opt are more obvious in this case, since the censoring rate is 5.27\%.
 

\section{Conclusions}
\label{sec5}
We develop an optimal subsampling strategy that can be used with massive survival data to overcome the computational limitations of the Cox proportional hazards model. It is proven theoretically that the proposed subsample estimator can approximate the estimator using the full dataset. In terms of its asymptotic properties, the developed estimator is consistent and asymptotically normal with the optimal variance. Further, numerical studies show that the proposed approach can reduce the computational burden and improve estimation efficiency. Using real-world data from a commercial bank in China and the National Cancer Institute, the superiority of the proposed method over other subsampling-based methods is verified.

This study can be extended in multiple ways in future research. First, we consider the optimal subsampling probabilities by estimating the regression parameter, which may be suboptimal for estimating the baseline hazard function. Hence, future studies could examine the optimal subsampling probabilities for the baseline hazard function. Second, the proposed method focuses on the cases of right-censored data with the Cox proportional hazards model. Future researchers could consider other survival models such as the transformation model and accelerated failure time model and use right-censored and left-truncated data.

\section*{Acknowledgements}
 This work was supported by the Fundamental Research Funds for the Central Universities, and the Research Funds of Renmin University of China (No.19XNB014).

\bibliographystyle{apalike}
\bibliography{Reference}

\begin{thebibliography}{}

\bibitem[Agrawal and Maheshwari, 2019]{Agrawal2019}
Agrawal, K. and Maheshwari, Y. (2019).
\newblock Efficacy of industry factors for corporate default prediction.
\newblock {\em IIMB Management Review}, 31(1):71--77.

\bibitem[Ai et~al., 2021]{ai2021optimal}
Ai, M., Yu, J., Zhang, H., and Wang, H. (2021).
\newblock Optimal subsampling algorithms for big data regressions.
\newblock {\em Statistica Sinica}, 31(2):749--772.

\bibitem[Bellotti and Crook, 2009]{bellotti2009credit}
Bellotti, T. and Crook, J. (2009).
\newblock Credit scoring with macroeconomic variables using survival analysis.
\newblock {\em Journal of the Operational Research Society}, 60(12):1699--1707.

\bibitem[Chen and Zhou, 2020]{chen2020}
Chen, L. and Zhou, Y. (2020).
\newblock Quantile regression in big data: A divide and conquer based strategy.
\newblock {\em Computational Statistics \& Data Analysis}, 144:106892.

\bibitem[Chen and Xie, 2014]{chen2014split}
Chen, X. and Xie, M. (2014).
\newblock A split-and-conquer approach for analysis of extraordinarily large
  data.
\newblock {\em Statistica Sinica}, pages 1655--1684.

\bibitem[Cheng et~al., 2020]{cheng2020information}
Cheng, Q., Wang, H., and Yang, M. (2020).
\newblock Information-based optimal subdata selection for big data logistic
  regression.
\newblock {\em Journal of Statistical Planning and Inference}, 209:112--122.

\bibitem[Cox, 1972]{cox1972}
Cox, D.~R. (1972).
\newblock Regression models and life-tables.
\newblock {\em Journal of the Royal Statistical Society: Series B
  (Methodological)}, 34(2):187--202.

\bibitem[Cox, 1975]{cox1975partial}
Cox, D.~R. (1975).
\newblock Partial likelihood.
\newblock {\em Biometrika}, 62(2):269--276.

\bibitem[Dai et~al., 2020]{dai2020}
Dai, W., Jiang, X., Bonomi, L., Li, Y., Xiong, H., and Ohno-Machado, L. (2020).
\newblock Verticox: Vertically distributed cox proportional hazards model using
  the alternating direction method of multipliers.
\newblock {\em IEEE Transactions on Knowledge and Data Engineering}.

\bibitem[Djeundje and Crook, 2019]{djeundje2019dynamic}
Djeundje, V.~B. and Crook, J. (2019).
\newblock Dynamic survival models with varying coefficients for credit risks.
\newblock {\em European Journal of Operational Research}, 275(1):319--333.

\bibitem[Dobriban and Sheng, 2021]{Dobriban2021}
Dobriban, E. and Sheng, Y. (2021).
\newblock Distributed linear regression by averaging.
\newblock {\em The Annals of Statistics}, 49(2):918--943.

\bibitem[Keret and Gorfine, 2020]{keret2020optimal}
Keret, N. and Gorfine, M. (2020).
\newblock Optimal cox regression subsampling procedure with rare events.
\newblock {\em arXiv preprint arXiv:2012.02122}.

\bibitem[Lin and Ying, 1994]{lin1994}
Lin, D.~Y. and Ying, Z. (1994).
\newblock Semiparametric analysis of the additive risk model.
\newblock {\em Biometrika}, 81(1):61--71.

\bibitem[Ma et~al., 2015]{ma2015statistical}
Ma, P., Mahoney, M.~W., and Yu, B. (2015).
\newblock A statistical perspective on algorithmic leveraging.
\newblock {\em Journal of Machine Learning Research}, 16:861--911.

\bibitem[Rosenberg et~al., 2005]{Rosenberg2005}
Rosenberg, J., Chia, Y.~L., and Plevritis, S. (2005).
\newblock The effect of age, race, tumor size, tumor grade, and disease stage
  on invasive ductal breast cancer survival in the us seer database.
\newblock {\em Breast Cancer Research and Treatment}, 89(1):47--54.

\bibitem[Sedgewick, 1977]{sedgewick1977analysis}
Sedgewick, R. (1977).
\newblock The analysis of quicksort programs.
\newblock {\em Acta Informatica}, 7(4):327--355.

\bibitem[Tarkhan and Simon, 2020]{tarkhan2020}
Tarkhan, A. and Simon, N. (2020).
\newblock Bigsurvsgd: Big survival data analysis via stochastic gradient
  descent.
\newblock {\em arXiv preprint arXiv:2003.00116}.

\bibitem[Wang, 2019]{wang2019more}
Wang, H. (2019).
\newblock More efficient estimation for logistic regression with optimal
  subsamples.
\newblock {\em Journal of Machine Learning Research}, 20.

\bibitem[Wang and Ma, 2021]{wang2021optimal}
Wang, H. and Ma, Y. (2021).
\newblock Optimal subsampling for quantile regression in big data.
\newblock {\em Biometrika}, 108(1):99--112.

\bibitem[Wang et~al., 2019]{wang2019information}
Wang, H., Yang, M., and Stufken, J. (2019).
\newblock Information-based optimal subdata selection for big data linear
  regression.
\newblock {\em Journal of the American Statistical Association},
  114(525):393--405.

\bibitem[Yuan et~al., 2018]{yuan2018}
Yuan, M., Tang, C.~Y., Hong, Y., and Yang, J. (2018).
\newblock Disentangling and assessing uncertainties in multiperiod corporate
  default risk predictions.
\newblock {\em The Annals of Applied Statistics}, 12(4):2587--2617.

\bibitem[Zhang et~al., 2020]{zhang2020}
Zhang, X., Ouyang, R., Liu, D., and Xu, L. (2020).
\newblock Determinants of corporate default risk in china: The role of
  financial constraints.
\newblock {\em Economic Modelling}, 92:87--98.

\bibitem[Zuo et~al., 2021]{zuo2021sampling}
Zuo, L., Zhang, H., Wang, H., and Liu, L. (2021).
\newblock Sampling-based estimation for massive survival data with additive
  hazards model.
\newblock {\em Statistics in Medicine}, 40(2):441--450.

\end{thebibliography}

\end{document}